\documentclass[eqsecnum,aps,preprint]{revtex4}

\usepackage{latexsym}
\usepackage{graphicx}
\usepackage{url}
\usepackage{amsmath}
\usepackage{epsf}

\newcommand{\ve}[1]{\mbox{\boldmath$#1$}}
\newcommand{\GGV}{\langle \Omega |:\frac{\alpha_s}{\pi} G^2: | \Omega \rangle} 
\newcommand{\GGM}{\langle \Omega |:\frac{\alpha_s}{\pi}\left(\frac{(vG)^2}{v^2} - \frac{G^2}{4} \right):| \Omega \rangle} 
\newcommand{\TrCD}{{\rm Tr}_{\rm C,D}}
\newcommand{\TrC}{{\rm Tr}_{\rm C}}
\newcommand{\TrD}{{\rm Tr}_{\rm D}}

\newcommand{\GGVren}{\langle \Omega | \frac{\alpha_s}{\pi}G^2 | \Omega \rangle} 
\newcommand{\GGMren}{\langle \Omega | \frac{\alpha_s}{\pi} \left(\frac{(vG)^2}{v^2} - \frac{G^2}{4} \right) | \Omega 
\rangle} 

\begin{document}

\title{In-medium operator product expansion for heavy-light-quark pseudoscalar mesons}

\author{\sc Sven Zschocke$^{\bf 1,2}$, Thomas Hilger$^{\bf 3,4}$, Burkhard K\"ampfer$^{\bf 3,4}$}
\affiliation{$^{\bf 1}$ TU Dresden, Institut f\"ur Planetare Geod\"asie, Lohrmann-Observatorium, D-01062 Dresden, Germany\\
$^{\bf 2}$ Dept. of Physics and Technology of the University of Bergen, Allegaten 55, N-5007 Bergen, Norway\\
$^{\bf 3}$ Helmholtz-Zentrum Dresden-Rossendorf, PF 510119, D-01314 Dresden, Germany\\ 
$^{\bf 4}$ TU Dresden, Institut f\"ur Theoretische Physik, D-01062 Dresden, Germany}

\begin{abstract} 
The operator product expansion (OPE) for heavy-light-quark pseudoscalar mesons (D-mesons and B-mesons) in medium is 
determined, both for a moving meson with respect to the surrounding medium as well 
as for a meson at rest.
First of all, the OPE is given in terms of normal-ordered operators up to mass dimension 5, and the mass of the 
heavy-quark and the mass of the light-quark are kept finite. The Wilson coefficients of such an expansion are 
infrared (IR) divergent in the limit of a vanishing light-quark mass. A consistent separation of scales 
necessitates an OPE in terms of non-normal-ordered operators, which implies operator mixing,  where the IR-divergences 
are absorbed into the operators. It is shown that the Wilson coefficients of such an expansion are IR-stable, and 
the limit of a vanishing light-quark mass is perfomed. Details of the major steps for the calculation of the Wilson 
coefficients are presented. By a comparison with previous results obtained by other theoretical groups we have found 
serious disagreements. 
\end{abstract}

\maketitle

\newpage

\tableofcontents

\newpage

%================================================
\section{Introduction} \label{sct:intro}

The discovery of the charm-quark in 1974 by detecting the $J/\Psi$-particle \cite{discovery1,discovery2} 
has completed the second quark-generation and 
was a manifest triumph of the quark-model. While $J/\Psi$ is a bound state of a charm-quark and an anti-charm quark, 
$J/\Psi = \overline{c}c$, so-called open-charmed mesons were discovered soon afterwards: 
$D^{+} = c \overline{d}, D^{0} = c \overline{u}$, and their corresponding anti-mesons 
$D^{-} = d \overline{c}, \overline{D}^{0} = u \overline{c}$. The open charmed mesons are much 
easier accessible experimentally, because the D-mesons are the lightest particles which contain a charm-quark 
and their lifetime is a few orders of magnitude larger than $J/\Psi$: the mass of $J/\Psi$ is $3096.9\;{\rm MeV}$ and the 
lifetime is $7.06 \times 10^{-21}\;{\rm sec}$, while, for instance, the mass of $D^{\pm}$-meson is $1869.6\;{\rm MeV}$ 
and the lifetime $1.04 \times 10^{-12}\;{\rm sec}$ \cite{Particle_Data_Book}. 

Whereas charmed mesons in vacuum were studied thoroughly ever since, nothing is known so far from experimental side about 
the properties of these mesons in medium. Nowadays, experiments are initiated to study charmed mesons embedded in a 
hadronic medium. In fact, the upcoming accelerator FAIR (Facility for Antiproton and Ion Research) 
at GSI (Gesellschaft f\"ur Schwer-Ionenforschung) in Darmstadt/Germany \cite{FAIR} offers the opportunity to study charmed 
mesons in dense nuclear matter. Especially, the CBM (Compressed Baryonic Matter) collaboration intends to study the 
near-threshold production of D-mesons and $J/\Psi$ in heavy-ion collisions, while the PANDA 
(anti-Proton ANihilation at DArmstadt) collaboration will focus on charm spectroscopy as well as on 
charmed mesons produced by anti-proton annihilation in nuclei.

This growing interest can also be motivated from in-medium modifications of K-mesons, that is an 
expected down-shift of $K^{-}$ and an up-shift of $K^{+}$ with increasing density \cite{K_Mesons}. There is a similarity between D-mesons 
and K-mesons in respect to their quark structure: $K^{-} = s \overline{u}$ corresponds to $D^0$, while 
$K^{+} = u \overline{s}$ corresponds to ${\overline D}^0$. 
Thus, we conclude the qualitative fact, that both strangeness and charm might be 
regarded as probes of the in-medium situation.  
Moreover, the expected in-medium modifications of D-mesons might have a considerable impact on normal $J/\Psi$ suppression,  
e.g. \cite{J_Psi_suppression}, and open-charm enhancement, e.g. \cite{open_charm_enhancement}, in heavy-ion collisions. 

Despite the upcomming experiments CBM and PANDA, theoretical investigations of charmed mesons, both in vacuum and in 
medium, are fairly rare. One reason is, that the well-established perturbation theory of Quantum 
Chromodynamics (QCD), the fundamental theory of strong interactions, cannot be applied, because the momentum transfer $Q$ 
among bound quarks is small $Q \sim \Lambda_{\rm QCD} \simeq 245 \, {\rm MeV}$ \cite{Lambda_QCD}. Thus, the running 
coupling constant of QCD $\alpha_s(Q)$ becomes large, and non-perturbative approaches are compelling in order to derive 
the physical properties of mesons from fundamental principles.

Especially, the infrared regime of QCD is complicated due to the still unresolved phenomenon of confinement. One aspect 
of thereof is the non-vanishing of normal-ordered products of quark and gluon field operators taken between the 
non-perturbative, so-called physical, groundstate $|{\rm vac} \rangle$ of QCD, e.g. the chiral quark-- and 
gluon--condensate: 
\begin{eqnarray}
\langle {\rm vac} | : \overline{q}^{\rm a}\, q^{\rm a} : | {\rm vac} \rangle \neq 0 \,,
\label{intro_1}
\\
\langle {\rm vac} | : G_{\mu \nu}^{A} \, G^{A\;\mu \nu} : | {\rm vac} \rangle \neq 0 \,.
\label{intro_2}
\end{eqnarray}

\noindent
Here, $: {\cal O} :$ denotes normal-ordering of a 
composite operator ${\cal O}$, $q^a$ is the quark field operator of color index $a=1,2,3$, and $G_{\mu \nu}^{A}$ is the
gluon field strengh tensor, where $\mu,\nu=0,1,2,3$ are Lorentz-indices and $A=1,...,8$ is the Gell-Mann index; troughout
the article the Einstein convention is used. 
Vacuum expectation values like (\ref{intro_1}) and (\ref{intro_2}) are called vacuum-condensates, a term which refers to the fact that the physical ground 
state of QCD is not "empty" but contains colorless and chargeless states of strongly bounded
(condensed) quarks and gluons. In contrast, in the perturbative vacuum $|0\rangle$ of QCD the normal-ordered condensates
vanish, e.g.: $\langle 0 | : \overline{q}^a\, q^a : | 0 \rangle = 0$ and 
$\langle 0 | : G_{\mu \nu}^{A} \, G^{A\;\mu \nu} : | 0 \rangle =0$. 

At finite baryonic densities the vacuum-condensates are generalized to 
in-medium-condensates, where the in-medium state is denoted by $| \Omega \rangle$, 
i.e. 
\begin{eqnarray}
\langle \Omega | : \overline{q}^{\rm a}\, q^{\rm a} : | \Omega \rangle \,,
\label{intro_3a}
\\
\langle \Omega | : G_{\mu \nu}^{A} \, G^{A\;\mu \nu} : | \Omega \rangle\,.
\label{intro_4a}
\end{eqnarray}

\noindent
The ground state of a hadronic medium can be approximated as a Fermi-gas of nucleons: 
$|\Omega\rangle = |{\rm vac}\rangle + \sum_k | {\rm N}(k)\rangle$. 
It is a formidable challenge to evaluate vacuum and in-medium condensates from first 
principles of QCD. So far, approximative solutions have been obtained (e.g. lattice gauge theory, Wilson loop expansion, 
instantons, etc.), but a comprehensive understanding of the confinement problem is far out of reach. 
Therefore, the question arises how fundamental QCD parameters may be related to the vacuum and 
in-medium properties of mesons.

One of the best methods among the non-perturbative approaches are the QCD Sum Rules (QSR), which link hadronic observables 
like mass, decay-width and coupling-constant of the hadron under consideration to  fundamental parameters of QCD. 
The QSR have first been developed for mesons in vacuum \cite{Shifman1,Shifman2,Shifman3}, and later been generalized to the case of mesons in 
medium \cite{Medium1,Medium2,Medium3,Medium4}, see also \cite{Zschocke1} for the case of finite density and temperature. 
The decisive advantage of this approach is, that QSR 
consider the existence of vacuum--condensates and in-medium--condensates as a given fact, instead of determining their numerical 
values by first principles of QCD. In this way, QSR are not concerend with the problem of confinement, but assume that 
confinement exists and they are neither concerned with the involved structure of the groundstate $| {\rm vac} \rangle$ nor the in-medium state $| \Omega \rangle$.

Thus, once a very few parameters, i.e. numerical values of condensates, are given, physical properties of a large variety
of hadrons can be predicted. The link between hadronic observables and fundamental parameters of QCD is given by a 
dispersion relation of n-point correlators which are non-local products of n operators: 
${\cal O}_1 (x_1) {\cal O}_2 (x_2) ... {\cal O}_n (x_n)$. Here, we consider so-called current-current 
correlators $j_1 (x)\,j_2 (y)$, which are two-point correlators. 

The analyticity of the current-current operator allows to relate the time-like 
region (hadronic part) with the space-like region (QCD part) of that operator. 
While the hadronic side of the dispersion relation is
parametrized by means of hadronic observables, the Wilson Operator Product Expansion (OPE) \cite{Wilson} is applied on the QCD side. In general, the 
product $j_1(x) j_2(y)$ diverges in 
the limit $x \rightarrow y$, e.g. \cite{Zuber,Muta,Pascual_Tarrach}. The Wilson OPE allows to
decompose the product of non-local operators in terms of a series of regular local operators ${\cal O}$ and so-called 
Wilson coefficients $C_k$ which are divergent for $x \rightarrow y$, i.e.
\begin{eqnarray} 
j_1(x)\,j_2(y) &=& \sum\limits_{k=0}^{\infty} C_k \left(\frac{x-y}{2}\right) {\cal O}_k \left(\frac{x+y}{2}\right). 
\label{intro_5} 
\end{eqnarray}

\noindent
The point of the OPE (\ref{intro_5}) is that the fields of the currents are separated into a hard and a soft part. The hard 
part is proportional to the unit operator and can be treated perturbatively, i.e. can be evaluated with respect to 
the perturbative vacuum $|0\rangle$. The spatial dependence of the soft part can be Taylor expanded and leads to local 
condensates in the medium state $|\Omega\rangle$.

In general, the currents of the correlator are interpolating fields, 
which carry the symmetries of the specific hadron: spin, isospin, 
parity, charge and the valence quark content. More 
specific, in case of D-mesons or B-mesons we need to analyze correlators of heavy-light-quark currents. OPE's of such 
heavy-light-quark current-current correlators have been evaluated since the early days of QCD sum rules, 
see pioneering investigations \cite{Aliev,Narison1,Reinders1,Reinders2,Reinders3,Broadhurst1,Broadhurst2,Generalis1}.

One of the main peculiarities within the calculation of the OPE for a pseudoscalar heavy-light-quark meson correlator is the
absorption of infrared mass divergences which occur in the Wilson coefficients. In order to render the OPE finite, these
divergences have to be absorbed into the condensates by introducing non-normal-ordered condensates, like 
\begin{eqnarray}
\langle \Omega |\overline{q}^{\rm a}\, q^{\rm a} | \Omega \rangle \,,
\label{intro_3b}
\\
\langle \Omega | G_{\mu \nu}^{A} \, G^{A\;\mu \nu} | \Omega \rangle\,,
\label{intro_4b}
\end{eqnarray}

\noindent
instead of normal-ordered ones. The unique mathematical scheme behind is the approach of operator mixing which allows for a 
consistent separation of scales. This problem has been worked out in detail in \cite{Generalis1,Jamin_Munz}, and accounts 
for a perturbative piece of the condensates, cf.~\cite{Chetyrkin2,Generalis1,Jamin_Munz,Broadhurst1} for a detailed 
discussion of the vacuum case. 

At non-vanishing baryonic densities the OPE differs from the vacuum case and additional mass singularities occur. In order 
to treat them consistently one has to find additional expressions which reproduce the vacuum limit but also render the 
in-medium OPE infrared stable. A crucial point is the mixing of different condensates under this procedure. The in-medium 
OPE for open charmed mesons has been evaluated in \cite{Hayashigaki1,Morath1,Morath2}. Unfortunately, the results and Wilson 
coefficients presented in these references significantly differ from each other. Moreover, it is not clear in which way  
the authors have dealt with the infrared mass singularities.  
However, the knowledge of the correct OPE is compelling for a reliable prediction of in-medium properties of D-mesons 
within the QSR approach. Recently, in \cite{D_Meson1,D_Meson2} an OPE for heavy-light currents in medium has been presented where the 
operator mixing and cancellation of infrared simgularities were correctly taken into account, but no further details of the involved 
evaluation of the OPE were given. In view of the progressing in-medium D-meson physics, both experimentally and theoretically, 
it is timely to present a transparent and thorough calculation of the OPE for D-mesons in matter and 
a comprehensive representation of the techniques which have to be applied. 

The pseudoscalar B-mesons are also heavy-light-quark systems and their quark structure is: 
$B^{+} = u \overline{b}$, $B^{-} = b \overline{u}$, $B^{0} = d \overline{b}$ and $\overline{B}^{0} = b \overline{d}$. 
So far, the only application of the QSR approach has been performed in \cite{D_Meson1} in order to determine in-medium 
modifications of B-mesons. Since the formalism of OPE and QSR can easily be extended from D-mesons to the case of B-mesons, 
we will incorporate these mesons in our investigation. 

The paper is organized as follows: The OPE in terms of normal-ordered condensates is calculated in Section \ref{sct:ope}.
Non-normal-ordered condensates and, therewith associated, the operator mixing and the absorption of infrared divergences 
is discussed in Section \ref{sct:abs_div}. The IR-limit, i.e. the limit of a vanishing light-quark mass, is 
considered in Section \ref{sct:Limit}. In Section \ref{sct:rest_frame} the OPE is given for the physical situation 
where the meson is at rest with respect to the surrounding medium. Furthermore, a Borel transformation is performed. 
A comparison with the results given in several publications is 
given in Section \ref{sct:Literature}. The summary can be found in Section \ref{sct:summary}. 

%======================================================

\section{OPE in terms of normal-ordered operators} \label{sct:ope}

\subsection{OPE with IR-divergent Wilson coefficients}

Let us introduce the current-current correlator for D-mesons, which is defined as Fourier transformation of 
the expectation value of the time-ordered products of two currents:
\begin{eqnarray} 
\Pi(q) &=& i \int d^4x \,e^{i q x} \langle \Omega |\,{\rm T} j (x) j^\dagger (0) \,| \Omega \rangle \,,
\label{eq:correlator}
\end{eqnarray} 

\noindent
where ${\rm T}$ denotes the Wick time-ordering, $q^{\mu} = (q_0, \ve{q})$ is the four-momentum of the D-meson, and 
$| \Omega \rangle$ is the groundstate of hadronic matter. Application of Wick's theorem to the correlator 
(\ref{eq:correlator}) naturally yields the OPE in terms of normal-ordered operators and can be written as 
\begin{eqnarray} 
\Pi(q) &=& \overline{C}_0 \,{\cal I} + \sum\limits_{i} \overline{C}_i \;\langle \Omega | : {\cal O}_i : | \Omega \rangle^{(0)} \,,
\label{OPE_IR_unstable}
\end{eqnarray} 

\noindent
where ${\cal I}$ is the unit operator, the notation $^{(0)}$ means tree-level matrix elements, and the bar denotes Wilson 
coefficients which correspond to tree-level matrix elements, cf. the OPE of heavy-light-quark currents 
in vacuum \cite{Jamin_Munz}. From now on we drop the label $^{(0)}$, i.e. normal-ordered condensates are always on tree-level throughout the work.  

As mentioned in the Introduction, the currents in (\ref{OPE_IR_unstable}) are constructed such that they contain the quantum numbers of the particle 
under consideration, i.e. spin, isospin, parity, charge and the valence quark 
content. Accordingly, for D-mesons they are given by 
\begin{eqnarray} 
j_{D^+} (x) &=& : i\,\overline{d}\,\gamma_5\,c : \;,
\label{eq:current_D1}
\\
j_{D^-} (x) &=& : i\,\overline{c}\,\gamma_5\,d : \;,
\label{eq:current_D2}
\\
j_{D^0} (x) &=& : i\,\overline{u}\,\gamma_5\,c : \;,
\label{eq:current_D3}
\\
j_{\overline{D}^0} (x) &=& : i\,\overline{c}\,\gamma_5\,u : \;.
\label{eq:current_D4}
\end{eqnarray} 

\noindent
Due to ${j}_{D^+}^{\dagger} = j_{D^-}$ and ${j}_{D^0}^{\dagger} = j_{\overline{D}^0}$ we obtain, by means of $j(x) = {\rm exp} (i P x)\,j(0)\,{\rm exp} (- i P x)$  where $P$ is the momentum operator, the relations 
$\Pi_{D^+} (q) = \Pi_{D^-} (-q)$ and $\Pi_{D^0} (q) = \Pi_{\overline{D}^0} (-q)$, respectively. Furthermore, in isospin symmetric nuclear matter a replacement 
$u \leftrightarrow d$ does not change these four 
correlators. Thus, it is sufficient 
to consider the current-current correlator (\ref{eq:correlator}) with the current operator of the $D^+$ meson given by 
Eq.~(\ref{eq:current_D1}), since the other three correlators $\Pi_{D^-}$, 
$\Pi_{D^0}$ and $\Pi_{\overline{D}^0}$, can easily be 
deduced from $\Pi_{D^+}$. 

Similarly, for the B-mesons the currents read:
\begin{eqnarray}
j_{B^+} (x) &=& : i\,\overline{b}\,\gamma_5\,u : \;,
\label{eq:current_B1}
\\
j_{B^-} (x) &=& : i\,\overline{u}\,\gamma_5\,b : \;,
\label{eq:current_B2} \\
j_{B^0} (x) &=& : i\,\overline{b}\,\gamma_5\,d :\;,
\label{eq:current_B3} \\
j_{\overline{B}^0} (x) &=& : i\,\overline{d}\,\gamma_5\,b : \;.
\label{eq:current_B4}
\end{eqnarray}

\noindent
Obviously, we have $j_{B^-}=j_{B^+}^{\dagger}$ and $j_{\overline{B}^0}=j_{B^0}^{\dagger}$ and 
$\Pi_{B^+} (q) = \Pi_{B^-} (-q)$ and $\Pi_{B^0} (q) = \Pi_{\overline{B}^0} (-q)$. Therefore, the OPE for the B-mesons can 
be deduced from the OPE of $B^+$. Furthermore, since the OPE of $B^-$ can be obtained from the OPE of $D^+$ by the replacements 
$u \rightarrow d$ and $b \rightarrow c$, we conclude that the OPE of all heavy-light-quark pseudoscalar currents can be deduced from 
the OPE of the $D^+$ meson; therefore, from now on we will drop the explicit 
notation $D^+$.

In this Section, we will determine the Wilson coefficients of (\ref{OPE_IR_unstable}). 
The OPE contains all operators up to mass dimension 5:
\begin{eqnarray}
{\cal I} &:& {\rm mass}\;{\rm dimension} \;0\;,
\label{Operator_0}
\\
\overline{q}_i^a \,q_j^b &:& {\rm mass}\;{\rm dimension} \;3\;,
\label{Operator_1}
\\
\overline{q}_i^c \,\overrightarrow{D}_{\mu}^{c a}\,q_j^b &:& {\rm mass}\;{\rm dimension} \;4\;,
\label{Operator_2}
\\
G_{\alpha\beta}^A\,G_{\mu\nu}^B &:& {\rm mass}\;{\rm dimension} \;4\;,
\label{Operator_3}
\\
\overline{q}_i^c \,\overrightarrow{D}_{\mu}^{c d}\,\overrightarrow{D}_{\nu}^{d a}\,q_j^b &:& {\rm mass}\;{\rm dimension}\;5\;,
\label{Operator_4}
\\
\overline{q}_i^c \sigma^{\mu\nu}\,{\cal G}_{\mu\nu}^{c a}\,q_j^b &:& {\rm mass}\;{\rm dimension} \; 5\;,
\label{Operator_5}
\end{eqnarray}

\noindent
where the flavor of the quark-fields are either charm or down quarks, 
the Dirac indices are denoted by $i,j = 1,2,3,4$, and the operators have to 
carry a colorless structure; for notation see Appendix \ref{sct:not}.  
In what follows, we will calculate all Wilson coefficients at first non-trivial 
order in powers of the QCD coupling constant, except for the unity operator where 
we include $\alpha_s$ corrections. 

As we will see in this Section, the Wilson coefficients of (\ref{OPE_IR_unstable}) are divergent in the limit of a vanishing 
light-quark mass $m_d \rightarrow 0$, which is a so-called IR-divergency. The reason for this divergence is that a proper 
factorization of short and long distance contributions in the OPE requires the calculation of matrix elements at the same 
order as the Wilson coefficients. This special issue will be the topic of the next Section. In this Section we will 
consider the OPE (\ref{OPE_IR_unstable}), i.e. in terms of normal-ordered operators.

By inserting the current (\ref{eq:current_D1}) into (\ref{eq:correlator}) and applying Wick's theorem to the current-current 
correlation function $\Pi(q)$ we obtain:
\begin{eqnarray} 
\Pi(q) &=& \Pi^{(0)} (q) + \Pi_d^{(2)} (q) + \Pi_c^{(2)} (q) +\Pi^{(4)} (q) \,,
\label{eq:Wick_Theorem_0}
\\
\nonumber\\
\Pi^{(0)} (q) &=& -i \int d^4x \, e^{iqx} \langle \Omega | : \TrCD \left(
\gamma_5\,S_d (0,x)\, \gamma_5 \,S_c (x,0) \right) : | \Omega \rangle\,,
\label{eq:Wick_Theorem_1}
\\
\Pi_d^{(2)} (q) &=& \int d^4x \, e^{iqx} \langle \Omega | : \overline{d} (x) \,\gamma_5\,S_c (x,0) \,\gamma_5\,d (0) 
: | \Omega \rangle\,,
\label{eq:Wick_Theorem_2_d}
\\
\Pi_c^{(2)} (q) &=& \int d^4x \, e^{iqx} \langle \Omega | : 
\overline{c} (0) \, \gamma_5\,S_d (0,x) \,\gamma_5\, c (x): | \Omega \rangle\,,
\label{eq:Wick_Theorem_2_c}
\\
\Pi^{(4)} (q) &=&  -i \int d^4x \, e^{iqx} 
\langle \Omega | : \overline{d} (x) \,\gamma_5\,c(x) \;\overline{c} (0) \,\gamma_5\,d(0) : | \Omega \rangle\,,
\label{eq:Wick_Theorem_3}
\end{eqnarray} 

\noindent
where the notation $\TrCD$ means trace over Dirac- and color-indices.  

The point of the whole OPE is that the quark and gluon fields are separated into a 
hard and a soft part. The hard part can be treated perturbatively. 
The soft part cannot be calculated in this way, 
but being soft one can Taylor expand the spatial dependence of the fields and 
relate it to local condensates. 

Accordingly, the term 
$\Pi^{(0)}$ is decomposed into a perturbative part, which is proportional to unit matrix and is treated 
by standard perturbation theory of QCD, and a gluonic part where the gluon fields are soft:
\begin{eqnarray}
\Pi^{(0)} (q) &=& \Pi^{\rm pert} (q) + \Pi^{(0)}_{G^2} (q) \;.
\label{Decomposition}
\end{eqnarray}

\noindent
The terms $\Pi^{(0)}_{G^2}$, $\Pi^{(2)}_d$, $\Pi^{(2)}_c$ and $\Pi^{(4)}$ describe the non-perturbative part of the 
correlator. The labels $^{(0)}$, $^{(2)}$ and $^{(4)}$ denote the number of non-contracted quark fields, 
i.e. the number of quarks which participate in the formation of a condensate. 
The term $\Pi^{(4)}$ concerns only the soft part of all operators. Thus there is no 
flow of hard momentum. The result of the integral in (\ref{eq:Wick_Theorem_3}) is 
proportional to Dirac's delta-function $\delta(q^2)$ and derivatives thereof, i.e. $\Pi^{(4)}$ vanishes 
except for $q^2 = 0$. Therefore at large $|q^2|$ (OPE) there is no contribution from 
this term. 

The quark propagator in a weak gluonic background field in coordinate space reads ($n$ is either $c$ or $d$)
\begin{eqnarray}
i S_n (x,y) &=& i S_n^{(0)} (x-y)
\nonumber\\
&& \hspace{-2.5cm} + \sum_{k=1}^\infty \int d^4z_1 \ldots d^4z_k\; 
i S_n^{(0)} (x-z_1) g_s \hat{{\cal A}}(z_1) i S_n^{(0)} (z_1-z_2) \ldots g_s 
\hat{{\cal A}}(z_k) i S_n^{(0)} (z_k-y)\,,
\nonumber\\
\label{eq:pert_quark_prop}
\end{eqnarray}

\noindent
where $S^{(0)}(x-y)$ denotes the free propagator and 'hat' a contraction with Dirac matrices
$\hat{{\cal A}} = \gamma_\mu {\cal A}^\mu$. For our investigation we need the 
quark propagator up to order $k=2$, given explicitly by Eqs.~(\ref{Quark_Propagator_10}) - (\ref{Quark_Propagator_20}). 

In what follows we will determine the Wilson coefficients of 
(\ref{eq:Wick_Theorem_1}) - (\ref{eq:Wick_Theorem_2_c}), while 
(\ref{eq:Wick_Theorem_3}) does not contribute at large $|q^2|$.

\subsection{The perturbative part $\Pi^{\rm pert}$}

The perturbative part is the Wilson coefficient of the unit operator (\ref{Operator_0}), and is given by 
\begin{eqnarray}
\Pi^{\rm pert}(q) &=& - i \int \frac{d^4 p}{(2\,\pi)^4} \; 
\TrCD \left( \gamma_5\,S_c^{(0)} (p) \,\gamma_5\,S_d^{(0)} (p-q)\right) + {\cal O} \left(\alpha_s\right).
\label{Im_Pert_0}
\end{eqnarray}

\begin{figure}
\begin{center}
\includegraphics{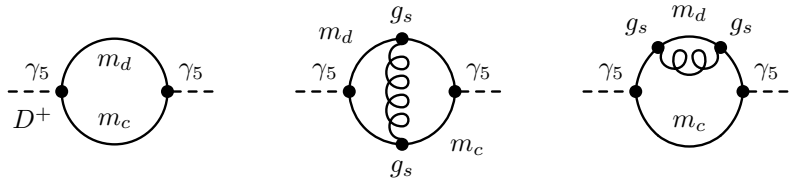}
\end{center}
\caption{Feynman diagrams for the perturbative part 
$\Pi^{\rm pert}(q)$ in Eq.~(\ref{Im_Pert_0}); the fourth diagram is not shown but can be obtained from the right one 
by the replacement $d \leftrightarrow c$. Solid lines
depict the free quark propagator, curly lines stand for contracted gluon fields, i.e. free gluon propagator,
and dashed lines denote the D-meson.
The left diagram is for the zeroth order ${\cal O} (\alpha_s^0)$, while the other
two diagrams are the first order ${\cal O} (\alpha_s)$ in perturbation theory of QCD.}
\label{Diagram_Perturbative}
\end{figure}

\noindent
Note, there is neither a normal-ordering nor an expectation value, since in 
perturbative QCD the propagator is a usual c-number. 
The first term in (\ref{Im_Pert_0}) corresponds 
to a one-loop Feynman diagram; see the left diagram 
in Fig.~\ref{Diagram_Perturbative}. The terms of order ${\cal O} \left(\alpha_s\right)$ are not given explicitly and correspond to two-loop Feynman 
diagrams, see the middle and the right diagram in Fig.~\ref{Diagram_Perturbative}. Their mathematical structure is almost identical 
to (\ref{Gluonic_Part_A}) - (\ref{Gluonic_Part_C}), but the two gluon fields are not soft anymore but contracted to 
a free gluon-field propagator; for equal quark mass these terms can be found in standard text books, 
e.g. \cite{Zuber,Pascual_Tarrach}. 

\noindent
The explicit solution of (\ref{Im_Pert_0}) is rather involved and is usually determined by means of the dispersion 
relation
\begin{eqnarray} 
\overline{C}_0 &\equiv& \Pi^{\rm pert}(q) 
= \frac{1}{\pi} \int\limits_{(m_c+m_d)^2}^{\infty} \,d s\;\frac{{\rm Im}\,\Pi^{\rm pert}(s)}{s - q^2}\,.
\label{Im_Pert_5}
\end{eqnarray} 

\noindent
The imaginary part of the perturbative term can be determined by means of the Cutkosky cutting rules \cite{Cutkosky1} (for 
a  didactical representation of the cutting rules including some examples see \cite{Cutkosky2}), and has been calculated at 
the very first time in \cite{Reinders2}, and later on several times by different authors, 
especially \cite{Broadhurst2,Aliev}; note that an error in the calculation of 
\cite{Reinders2} has been corrected by \cite{Broadhurst2}. In the $\overline{MS}$-scheme \cite{MS_bar}, the 
perturbative part is IR-convergent \cite{Tkachov1}. In accordance with the whole approach, we present our result for $m_d=0$ 
in Feynman gauge, which agrees with the findings of \cite{Aliev,Broadhurst2}:
\begin{eqnarray} 
{\rm Im}\,\Pi^{\rm pert}(s) &=& 
\frac{3}{8\,\pi}\,\frac{\left(s - m_c^2\right)^2}{s} 
\nonumber\\
&& \hspace{-2.0cm} +  \frac{\alpha_s}{2\,\pi^2}\,\frac{\left(s - m_c^2\right)^2}{s} \, 
\Bigg[ 
\frac{9}{4} + 2\,{\rm Li}_2 \left(\frac{m_c^2}{s}\right)
 + \ln \left(\frac{s}{m_c^2}\right) \ln \left(\frac{s}{s - m_c^2}\right) 
\nonumber\\
&& \hspace{-2.0cm} + \frac{3}{2}\,\ln \left(\frac{m_c^2}{s - m_c^2}\right) + \ln \left(\frac{s}{s - m_c^2}\right) + \frac{m_c^2}{s} \,
\ln \left(\frac{s - m_c^2}{m_c^2}\right) + \frac{m_c^2}{s - m_c^2} \,\ln \left(\frac{s}{m_c^2}\right)
\Bigg]\,.
\nonumber\\
\label{Im_Pert_10}
\end{eqnarray} 

\noindent
Here, ${\rm Li}_2 (x) = - \int\limits_0^x dt\;t^{-1}\,\ln (1-t)$ is the Spence function. We note, that the perturbative part must be symmetric in exchanging c-quark and d-quark. However, since we 
have presented the perturbative part in the IR-limit $m_d=0$, the given solution (\ref{Im_Pert_10}) is not symmetric 
anymore. 

We will not finish this paragraph without a final remark about the dispersion relation (\ref{Im_Pert_5}). 
According to (\ref{Im_Pert_10}), the dispersion relation (\ref{Im_Pert_5}) gives an infinite result in ultraviolet regions 
of integration, i.e. one might prefer a twice-subtracted dispersion relation instead of (\ref{Im_Pert_5}). Such a 
subtraction scheme yields polynomials in the external momentum $q$. However, after a Borel transformation has been 
performed, all polynomials disappear, that means a Borel transformation of a twice-subtracted dispersion relation is 
identical to a Borel transformation of (\ref{Im_Pert_5}).

\subsection{The gluonic part $\Pi^{(0)}_{G^2}$}

\begin{figure}
\begin{center}
\includegraphics{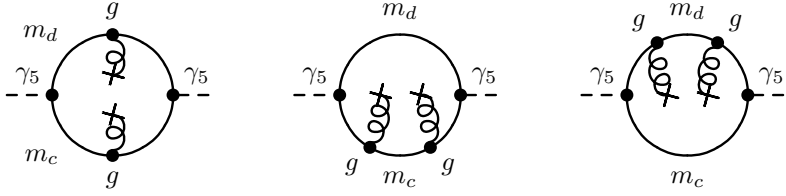}
\end{center}
\caption{Feynman diagrams corresponding to $\Pi^{G^{2,A}}$, $\Pi^{G^{2,B}}$ and $\Pi^{G^{2,C}}$ in 
Eqs.~(\ref{Gluonic_Part_A}), (\ref{Gluonic_Part_B}) and (\ref{Gluonic_Part_C}) from left to right. 
Solid lines depict the free quark propagator, dashed lines denote the D-meson,  
and curly lines stand for the soft gluon fields; the crosses symbolize the 
creation or annihilation of a gluon by virtual particles.}
\label{Diagram_Gluon}
\end{figure}

Let us turn to the gluonic part, Eqs.~(\ref{eq:Wick_Theorem_1}) and (\ref{Decomposition}), 
which is the result of inserting the next-to-leading order propagator with 
the lowest order term of Eq.~(\ref{eq:gluon}), and accounts for pure gluon condensates (\ref{Operator_3}). 
The expressions up to order ${\cal O} \left(\alpha_s\right)$ are given by:
\begin{eqnarray}
\Pi^{(0)}_{G^2}(q) &=& \Pi^{G^{2,A}} (q) + \Pi^{G^{2,B}} (q) + \Pi^{G^{2,C}} (q) \,,
\label{Gluonic_Part}
\\
\nonumber\\
\Pi^{G^{2,A}} (q) &=& - i  \int \frac{d^4 p}{\left(2 \pi\right)^4}
\langle \Omega | : \TrCD \left( \gamma_5\,S_c^{(1)} (p) \,\gamma_5\,S_d^{(1)} (p-q) \,\right):|\Omega\rangle\,,
\label{Gluonic_Part_A}
\\
\Pi^{G^{2,B}} (q) &=& - i  \int \frac{d^4 p}{\left(2 \pi\right)^4}
\langle \Omega | : \TrCD \left( \gamma_5\,S_c^{(2)} (p) \,\gamma_5\,S_d^{(0)} (p-q) \,\right) :|\Omega\rangle\,,
\label{Gluonic_Part_B}
\\
\Pi^{G^{2,C}} (q) &=& - i  \int \frac{d^4 p}{\left(2 \pi\right)^4}
\langle \Omega | : \TrCD \left( \gamma_5\,S_c^{(0)} (p+q) \,\gamma_5\,S_d^{(2)} (p) \,\right) :|\Omega\rangle\,,
\label{Gluonic_Part_C}
\end{eqnarray}

\noindent
where the quark propagators are given by Eqs.~(\ref{Quark_Propagator_10}) - (\ref{Quark_Propagator_20}).
These expressions correspond to the Feynman diagrams shown in Fig.~\ref{Diagram_Gluon}.

\noindent
First we note, that by inserting the expressions 
(\ref{Quark_Propagator_10}) - (\ref{Quark_Propagator_20}) into (\ref{Gluonic_Part_A}) - (\ref{Gluonic_Part_C}) we recognize 
that these three integrals are ultraviolet finite. We also note, that inserting higher orders of the quark 
propagator, e.g. $S_d^{(3)}$, or the gluon fields would lead to higher dimensional gluon 
condensates, such as $\langle : G^3 : \rangle$, or, by usage of the equations of motion, to quark and mixed quark-gluon 
condensates. These terms are either of higher mass dimension or of higher orders in $\alpha_s$, so we omit such terms. 

At this stage of our investigation we would like to mention that the calculation of the gluonic part $\Pi^{(0)}_{G^2}$ is 
performed at the one-loop level and leads to cumbersome expressions. Some details of the evaluation are therefore shifted 
to Appendix \ref{Gluonic_Terms}, where we have described in detail the techniques by means of which we obtain the 
following final result for the terms (\ref{Gluonic_Part_A}) - (\ref{Gluonic_Part_C}): 
\begin{eqnarray} 
\Pi^{(0)}_{G^2}(q) &=& \langle \Omega | : \frac{\alpha_s}{\pi}\;G^2: | \Omega \rangle 
\left(- \frac{1}{12} \frac{m_c}{m_d} \frac{1}{q^2 - m_c^2}
- \frac{1}{24} \frac{q^2}{(q^2 - m_c^2)^2} \right)
\nonumber\\
&& \hspace{-1.7cm} 
+ \langle \Omega | : \frac{\alpha_s}{\pi}\; \left(\frac{\left(v\,G\right)^2}{v^2} - \frac{G^2}{4}\right) : | \Omega \rangle 
\left( q^2 - 4\frac{(vq)^2}{v^2} \right)
\nonumber\\
&& \hspace{-1.7cm} 
\times \Bigg(- \frac{1}{6}\,\frac{1}{(q^2 - m_c^2)^2} 
- \frac{1}{9} \,\frac{1}{(q^2 - m_c^2)^2}\,\ln \left(\frac{m_d^2}{m_c^2}\right) - \frac{2}{9}\,\frac{1}{(q^2 - m_c^2)^2}\,
\ln \left( - \frac{m_c^2}{q^2 - m_c^2} \right)\Bigg).
\nonumber\\
\label{Pi_G2}
\end{eqnarray} 

\noindent
Here, $G^2 = G_{\mu\nu}^A\,G^{A\;\mu\nu}$, and $v^{\mu} = (1,\ve{v})$ is the four-velocity of the surrounding medium. 
The first line in (\ref{Pi_G2}) is the scalar contribution which does not vanish in vacuum, while the other term is a 
medium-specific condensate and vanishes at zero density. One immediately observes an IR-divergergent term 
$\propto m_d^{-1}$ known from the vacuum OPE of D-mesons, while in medium there is an additional logarithmic IR-singularity 
$\propto \ln m_d^2$. These both IR singularities appear because in the corresponding diagram (right diagram in 
Fig.~\ref{Diagram_Gluon}) there are three light-quark propagators with the very same momentum, while in the other diagrams 
there are only one or two. 

We note the symmetry of (\ref{Gluonic_Part_A}) - (\ref{Gluonic_Part_C}) in 
exchanging the charm and down quark. However, since we have to perform the IR-limit
$m_d \rightarrow 0$ after operator mixing, we have taken into account the leading IR-divergent terms only, that means
$m_d^{-1}$ and $\propto \ln m_d^2$. Hence, the solution (\ref{Pi_G2}) is
no longer symmetric in exchanging down and charm quark. In Appendix \ref{Gluonic_Terms} further details 
are given about how the needed expansions in terms of a small d-quark mass destroy this kind of symmetry.

Furthermore, terms of the form $\ln q^2$ and $\ln m_d^2$ occur simultaneously and 
cannot be made small at the same time for $- q^2 \gg m_d^2$. However, as has been found in \cite{Tkachov1}, 
they are remnants of the large distance behavior, i.e. they originate from the small momentum contribution to the loop 
integrals. Their occurrence breaks the neat separation of scales, which is a necessary feature of every OPE, and must 
therefore be absorbed into the condensates. In \cite{Tkachov1} the author argues that these logarithms do not occur when 
the Wilson coefficients are calculated within a minimal subtraction scheme. Moreover, they can be absorbed into the 
condensates if one reexpresses normal-ordered condensates, which naturally emerge if one applies Wick's theorem to (\ref{eq:correlator}), by so-called non-normal-ordered ones. This procedure has been known for a 
long time in vacuum, cf. \cite{Chetyrkin1,Chetyrkin2,Jamin_Munz,Narison1,Narison2} and references therein, 
although an explicit formula could not be found by us. In 
Section \ref{sct:abs_div} we introduce a formula which relates normal-ordered and 
non-normal-ordered condensates in matter. 

\subsection{The term $\Pi_d^{(2)}(q)$}

Let us now consider the term (\ref{eq:Wick_Theorem_2_d}), which leads to condensates which contain down-quarks.  
After expanding the light-quark fields and performing the Fourier transformation we obtain the expression
\begin{eqnarray} 
\Pi_d^{(2)}(q) &=& \sum_{k=0}^\infty \frac{(-i)^k}{k!} 
\langle \Omega | : \left( \bar{d}_i \overleftarrow{D}_{\alpha_1}\ldots\overleftarrow{D}_{\alpha_k} \right)^a 
\left( \gamma_5 \partial^{\alpha_1}\ldots\partial^{\alpha_k} S_c(q) \gamma_5 \right)^{ij}_{ab} d_j^b : |\Omega \rangle \,. 
\nonumber\\
\label{eq:2q} 
\end{eqnarray} 

\begin{figure}
\begin{center}
\includegraphics{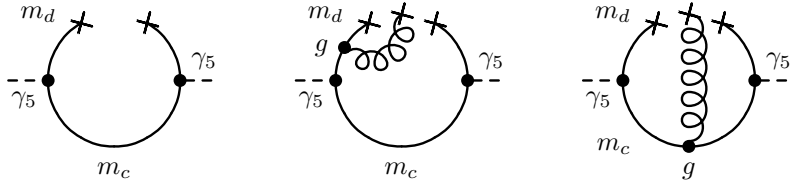}
\end{center}
\caption{The first three Feynman diagrams corresponding to $\Pi^{(2)}_d$ in 
Eq.~(\ref{eq:2q}). Solid lines depict the free quark propagator, 
dashed lines denote the D-meson, and curly lines stand for soft gluon fields; 
the crosses symbolize the creation or annihilation of either a gluon or a quark by 
virtual particles. The left diagram corresponds to the quark condensate 
$\langle :\overline{d} d : \rangle$, while the middle and right diagrams correspond 
to the mixed quark-gluon condensate $\langle :\bar{d}g\sigma{\cal G}d:\rangle$.}
\label{Diagram_Quark}
\end{figure}

\noindent
The quark fields and their covariant derivatives have to be calculated at the origin, i.e. $x=0$. From here we can go to 
higher quark field derivatives or to higher orders in the perturbative propagator or to higher orders in the gluon field, 
which enters through the perturbative quark propagator. The quark fields are of mass dimension $3/2$. 
Each covariant derivative and the gluon fields $\tilde{A}_\mu$ enlarge the mass dimension by one unit. Thus working in 
lowest order of the gluon fields the following terms have to be considered up to mass dimension 5:
\begin{eqnarray}
\Pi_d^{(2)}(q) &=&  \Pi_{d\,,\,A}^{(2)}(q) + \Pi_{d\,,\,B}^{(2)}(q) +\Pi_{d\,,\,C}^{(2)}(q) +\Pi_{d\,,\,D}^{(2)}(q) \,,
\label{second_term}
\end{eqnarray}

\noindent
where the individual contributions are 
\begin{eqnarray}
\Pi_{d\,,\,A}^{(2)}(q) &=& \langle \Omega | : \bar{d}_i^a \left( \gamma_5 S_c^{(0)}(q) \gamma_5 \right)^{ij}_{ab} d_j^b 
: | \Omega \rangle\,,
\label{before_projection_A}
\\
\nonumber\\          
\Pi_{d\,,\,B}^{(2)}(q) &=& - i \langle \Omega | : \left( \bar{d} \, \overleftarrow{D}_{\mu} \right)_i^a 
\left( \gamma_5 \partial^\mu S_c^{(0)}(q) \gamma_5 \right)^{ij}_{ab} d_j^b : | \Omega \rangle\,,
\label{before_projection_B}
\\
\nonumber\\
\Pi_{d\,,\,C}^{(2)}(q) &=& - \frac{1}{2} \langle \Omega | : \left( \bar{d} \, \overleftarrow{D}_{\mu} 
\overleftarrow{D}_{\nu} \right)_i^a \left( \gamma_5 \partial^\mu \partial^\nu S_c^{(0)}(q) \gamma_5 \right)^{ij}_{ab} 
d_j^b : | \Omega \rangle\,,
\label{before_projection_C}
\\
\nonumber\\
\Pi_{d\,,\,D}^{(2)}(q) &=&  \langle \Omega | : \bar{d}_i^a \left( \gamma_5 S_c^{(1)}(q) \gamma_5 \right)^{ij}_{ab} d_j^b : 
| \Omega \rangle\,,
\label{before_projection_D}
\end{eqnarray}

\noindent
where we note that the term (\ref{before_projection_B}) is not present in vacuum. 
In Fig.~\ref{Diagram_Quark} three corresponding Feynman diagrams are shown, but there is 
actually no one-to-one correspondence for the expressions (\ref{before_projection_A}) - (\ref{before_projection_D}) and Feynman diagrams.  
It becomes obvious, that the Wilson coefficients can be obtained on tree-level, in contrast to 
Wilson coefficients for gluon condensate which are obtained on one-loop level.

\noindent 
The calculations on tree-level are straightforward, so we just present the result: 
\begin{eqnarray}
\Pi_d^{(2)}(q) &=&
\langle \Omega | : \bar{d}d : | \Omega \rangle \frac{m_c}{q^2-m_c^2}
- \langle \Omega | : \bar{d} \,i\,\overrightarrow{D}_{\mu} d : | \Omega \rangle \frac{2\,m_c\,q^{\mu}}{(q^2-m_c^2)^2}
\nonumber\\
&& \hspace{-1.0cm} - \langle \Omega | : \bar{d} \gamma_{\lambda} d : | \Omega \rangle \frac{q^{\lambda}}{q^2-m_c^2}
+ \langle \Omega | : \bar{d} \gamma_{\lambda} \,i\,\overrightarrow{D}_{\mu} d : | \Omega \rangle
\left(\frac{2\,q^{\mu}\,q^{\lambda}}{(q^2-m_c^2)^2} - \frac{g^{\mu\lambda}}{q^2-m_c^2} \right)
\nonumber\\
&& \hspace{-1.0cm} + \langle \Omega | : \bar{d} \, \overrightarrow{D}_{\mu} \overrightarrow{D}_{\nu} d : | \Omega \rangle
\left( \frac{m_c\,g^{\mu\nu}}{(q^2-m_c^2)^2} - \frac{4\,m_c\,q^{\mu}\,q^{\nu}}{(q^2-m_c^2)^3} \right)
\nonumber\\
&& \hspace{-1.0cm} + \langle \Omega | : \bar{d} \gamma_{\lambda} \overrightarrow{D}_{\mu} \overrightarrow{D}_{\nu} d : | \Omega \rangle
\left( \frac{4\,q^{\mu}\,q^{\nu}\,q^{\lambda}}{(q^2-m_c^2)^3}
- \frac{q^{\mu}\,g^{\nu\lambda} + q^{\nu}\,g^{\mu\lambda} + q^{\lambda}\,g^{\mu\nu}}{(q^2-m_c^2)^2} \right)
\nonumber\\
&& \hspace{-1.0cm} - \langle \Omega | : \bar{d}\,g_s\,\sigma\,{\cal G}\, d : | \Omega \rangle
\frac{m_c}{2}\,\frac{1}{(q^2-m_c^2)^2} 
+ \langle \Omega | : \bar{d} g_s \gamma_5 \gamma_\alpha {\cal G}_{\mu\nu} d : | \Omega \rangle
\frac{1}{2} \frac{\epsilon^{\alpha\mu\nu\kappa}\,q_\kappa}{(q^2-m_c^2)^2} \;.
\nonumber\\
\label{before_Lorentz_projection_d}
\end{eqnarray}

\noindent
At this point of the evaluation we have to keep terms $\propto m_d$ (such terms appear when applying the equation of 
motion in some coondensates, e.g. in fourth condensate the term
proportional to metric tensor $g^{\mu\lambda}$) because they will be necessary to cancel the infrared 
divergences or give finite contributions to the correlator. The limit $m_d \to 0$ will be taken in the overnext 
Section after absorption of these infrared divergences.

\subsection{The term $\Pi_c^{(2)}(q)$}

Now we consider the term (\ref{eq:Wick_Theorem_2_c}) which leads to charmed condensates. 
In order to show the similarity with the expression $\Pi_d^{(2)}(q)$ obtained in the previous Subsection, 
we will keep the light-quark mass finite $m_d \neq 0$. 
The calculation of $\Pi_c^{(2)}(q)$ implies a tree-level evaluation, and we just present the result:  
\begin{eqnarray}
\Pi_c^{(2)}(q) &=&
\langle \Omega | : \bar{c}c : | \Omega \rangle \frac{m_d}{q^2-m_d^2}
- \langle \Omega | : \bar{c} \,i\,\overrightarrow{D}_{\mu} c : | \Omega \rangle \frac{2\,m_d\,q^{\mu}}{(q^2-m_d^2)^2}
\nonumber\\
&& \hspace{-1.0cm} + \langle \Omega | : \bar{c} \gamma_{\lambda} c : | \Omega \rangle \frac{q^{\lambda}}{q^2-m_d^2}
- \langle \Omega | : \bar{c} \gamma_{\lambda} \,i\,\overrightarrow{D}_{\mu} c : | \Omega \rangle
\left(\frac{2\,q^{\mu}\,q^{\lambda}}{(q^2-m_d^2)^2} - \frac{g^{\mu\lambda}}{q^2-m_d^2} \right)
\nonumber\\
&& \hspace{-1.0cm} + \langle \Omega | : \bar{c} \, \overrightarrow{D}_{\mu} \overrightarrow{D}_{\nu} c : | \Omega \rangle
\left( \frac{m_d\,g^{\mu\nu}}{(q^2-m_d^2)^2} - \frac{4\,m_d\,q^{\mu}\,q^{\nu}}{(q^2-m_d^2)^3} \right)
\nonumber\\
&& \hspace{-1.0cm} - \langle \Omega | : \bar{c} \gamma_{\lambda} \overrightarrow{D}_{\mu} \overrightarrow{D}_{\nu} c : | \Omega \rangle
\left( \frac{4\,q^{\mu}\,q^{\nu}\,q^{\lambda}}{(q^2-m_d^2)^3}
- \frac{q^{\mu}\,g^{\nu\lambda} + q^{\nu}\,g^{\mu\lambda} + q^{\lambda}\,g^{\mu\nu}}{(q^2-m_d^2)^2} \right)
\nonumber\\
&& \hspace{-1.0cm} - \langle \Omega | : \bar{c}\,g_s\,\sigma\,{\cal G}\,c : | \Omega \rangle
\frac{m_d}{2}\,\frac{1}{(q^2-m_d^2)^2} 
- \langle \Omega | : \bar{c} g_s \gamma_5 \gamma_\alpha {\cal G}_{\mu\nu} c : | \Omega \rangle
\frac{1}{2} \frac{\epsilon^{\alpha\mu\nu\kappa}\,q_\kappa}{(q^2-m_d^2)^2} \;.
\nonumber\\
\label{before_Lorentz_projection_c}
\end{eqnarray}

\noindent
Obviously, there are no IR-divergent terms due to charmed condensates. 
In our study we will neglect all charmed condensates (\ref{before_Lorentz_projection_c}). This assumption 
can be justified as follows. According to the approximation of the in-medium state  
$|\Omega \rangle = | {\rm vac} \rangle + \sum_k | {\rm N} (k) \rangle$, we can 
split the term (\ref{before_Lorentz_projection_c}) into a vacuum part and a 
nucleon part: $\Pi^{(2)}_c = \Pi^{(2)\,{\rm vac}}_c + \Pi^{(2)\,{\rm N}}_c$. 

At first, let us consider the vacuum part: 
$\Pi^{(2)\,{\rm vac}}_c$. In vacuum, the only non-vanishing term in the 
limit $m_d \rightarrow 0$ reads:
\begin{eqnarray}
\Pi_c^{(2)\,{\rm vac}}(q) &=&
- \langle {\rm vac} | : \bar{c} \gamma_{\lambda} \,i\,\overrightarrow{D}_{\mu} c : | 
{\rm vac} \rangle \left(\frac{2\,q^{\mu}\,q^{\lambda}}{q^4} 
- \frac{g^{\mu\lambda}}{q^2} \right). 
\label{before_Lorentz_projection_c_vacuum_1}
\end{eqnarray}

\noindent
We insert the operator mixing relation (\ref{physical_condensates_2c}) for the vacuum situation, and obtain 
the expression $\Pi_c^{(2)\,{\rm vac}}$ in terms of non-normal ordered condensates
\begin{eqnarray}
\Pi_c^{(2)\,{\rm vac}}(q) &=&
\frac{1}{2}\,\frac{m_c}{q^2}\,\langle {\rm vac} | \bar{c} c | {\rm vac} \rangle
+ \frac{1}{24\,q^2}\,\langle {\rm vac} | \frac{\alpha_s}{\pi} G^2 | {\rm vac} \rangle
- \frac{3}{8}\,\frac{m_c^4}{q^2}\, \left(\ln \frac{\mu^2}{m_c^2} + 1\right),
\nonumber\\
\label{before_Lorentz_projection_c_vacuum_2}
\end{eqnarray}

\noindent
where we also have used the projection of the Lorentz structure for the vacuum case: 
$\langle {\rm vac} | \bar{c} \gamma_{\lambda} \,i\,\overrightarrow{D}_{\mu} c | {\rm vac} \rangle =
\displaystyle \frac{m_c}{4}\,g_{\lambda\mu}\,\langle {\rm vac} | \overline{c}c | {\rm vac} \rangle$; 
cf. Eq.~(\ref{projection2a}) for the corresponding in-medium projection. 

Now we apply the heavy-quark mass (HQM) expansion, which is an expansion of a heavy quark condensate 
in inverse powers of the heavy quark mass, and for the scalar charmed condensate 
in vacuum it is given by \cite{Grozin_Pinelis,Gr94,Broadhurst3}:
\begin{eqnarray}
\langle  {\rm vac} | \overline{c} c | {\rm vac} \rangle &=& \frac{3}{4} m_c^3\,\left(\ln \frac{\mu^2}{m_c^2} + 1\right) 
- \frac{1}{12 m_c}\,\langle {\rm vac} | \frac{\alpha_s}{\pi} G^2 | {\rm vac} \rangle 
+ {\cal O} \left(m_c^{-3}\right).
\label{HQE_1}
\end{eqnarray}

\noindent
The idea behind the HQM expansion is, that the interaction of the heavy quarks with the QCD vacuum mainly happens via gluon 
interactions, because the quark itself is too heavy to couple directly to a condensate. Hence, the heavy quark condensates 
are expressed in terms of gluon condensates. Obviously, by inserting (\ref{HQE_1}) into 
(\ref{before_Lorentz_projection_c_vacuum_2}) we recognize that both the additional gluon condensate and the logarithmic 
term which appear in (\ref{before_Lorentz_projection_c_vacuum_2}) are exactly cancelled by the HQM expansion. 
This statement of cancellation can also be found e.g. in \cite{Reinders1}. 

Let us now consider the nucleon part: $\Pi^{(2)\,{\rm N}}_c$. The charmed condensates in a nucleon are negligibly 
small \cite{charm_nucleon1}, since they contribute in nucleon only via virtual effects; in this respect, to generate an 
easily interpreted image we recall the Tamm-Dancoff approach 
\cite{Tamm_Dancoff1,Tamm_Dancoff2,Tamm_Dancoff2a,Tamm_Dancoff2b, Tamm_Dancoff3} where the 
nucleon consists of a valence quark core of up and down quarks accompanied by a cloud of virtual mesons which accounts for 
the virtual sea quarks (e.g. charm quarks) and gluons. Thus, we can neglect the in-medium charmed condensates.  
Finally we note, that it is almost certain for us that the 
described cancellation of charmed condensates in vacuum via HQM expansion can 
be generalized to the in-medium charmed condensates, but a detailed proof
needs special care and would spoil the intention of our paper.

In summary of this Section, the complete OPE up to operators of mass dimension 5 and up to the first non-trivial order in 
the coupling constant in terms of normal-ordered operators is given by Eq.~(\ref{eq:Wick_Theorem_0}), where 
(\ref{eq:Wick_Theorem_1}) is given by Eqs.~(\ref{Im_Pert_5}) - (\ref{Pi_G2}), and (\ref{eq:Wick_Theorem_2_d}) is given by 
Eq.~(\ref{before_Lorentz_projection_d}); the term (\ref{eq:Wick_Theorem_2_c}) is given by
Eq.~(\ref{before_Lorentz_projection_c}) but will be neglected in our 
further investigation. Finally, the term (\ref{eq:Wick_Theorem_3})  
does not contribute at large $|q^2|$.

%================================================
\section{OPE in terms of non-normal-ordered operators} \label{sct:abs_div} 

\subsection{Absorption of IR divergences}

In order to perform a consistent separation of scales, all the infrared divergences have to be absorbed into the 
condensates, which means that the coefficient functions are only determined by the short distance behavior, while the 
non-perturbative effects are encoded in the condensates. In \cite{Ch82,Tkachov2,Smith_DeVries} it has been shown that the 
Wilson coefficients are polynomial functions of the mass only when they are calculated in a minimal subtraction scheme. 
In contrast, normal-ordering is not a minimal subtraction scheme. Especially, by introducing normal-ordered condensates 
the IR-divergent terms appear explicitly in the Wilson coefficients. For a small quark mass these IR-divergent terms 
are remnants of the long-distance part of the correlator and have to be absorbed into the 
condensates since they basically contain the long-distance part of the OPE. 
Therefore, we have to 
express the normal-ordered tree-level condensates $\langle :{\cal O}: \rangle^{(0)}$ in terms of 
non-normal-ordered condensates on one-loop level $\langle {\cal O} \rangle^{(1)}$. That means, the OPE in 
Eq.~(\ref{OPE_IR_unstable}) is reexpressed in terms of non-normal-ordered operators: 
\begin{eqnarray} 
\Pi(q) &=& C_0\,{\cal I} + \sum\limits_{i} C_i\;\langle \Omega | {\cal O}_i | \Omega \rangle^{(1)} \,.
\label{OPE_IR_stable}
\end{eqnarray} 

\noindent
In the expansion (\ref{OPE_IR_stable}) a proper separation of short and long distance contributions has been performed. 
Especially, the Wilson coefficients will be finite in the limit of a vanishing light-quark mass $m_d \rightarrow 0$, that 
means IR-stable. The Wilson coefficients in (\ref{OPE_IR_stable}) have no bar, which indicates that they are 
Wilson coefficients of non-normal-ordered operators on one-loop level. 
In consistency with (\ref{OPE_IR_unstable}), we have to take into account all operators up to mass dimension 5. 

\subsection{Operator mixing}

In order to express the normal-ordered condensates by physical condensates, we note Wick's theorem for an equal-time 
operator product of two quark fields. In doing so we partly follow \cite{Gr94} to get
\begin{eqnarray}
{\rm T}\,\bar{q}(x) \,{\cal O} \left[\overrightarrow{D}_\mu\right]\,q(y) &=&  : \bar{q}(x) \, 
{\cal O}\left[\overrightarrow{D}_\mu\right] \, q(y) : 
\; \, - \,i\; : \TrCD \left( {\cal O}\left[\overrightarrow{D}_\mu\right] S(y,x) 
\right) : \,.
\nonumber\\
\label{HQE_5}
\end{eqnarray}

\noindent
Here, ${\cal O}\left[\overrightarrow{D}_\mu\right]$ denotes a function of covariant derivatives in respect to $y$, 
gluon fields, and Dirac structures. Now we set $x=0$ in the expression (\ref{HQE_5}) and insert a Fourier representation 
of the quark propagator:
\begin{eqnarray}
{\rm T}\,\bar{q}(0) \,{\cal O} \left[\overrightarrow{D}_\mu\right]\,q(y) &=&  : \bar{q}(0) \, {\cal O}\left[\overrightarrow{D}_\mu\right] \, q(y) :
\nonumber\\
&& \hspace{-1.0cm} - i \int \frac{d^4 p}{\left(2\,\pi\right)^4} {\rm e}^{-i\,p\,y}\;
: \TrCD \left( {\cal O} \left[ -i\,p_\mu - i \tilde{A}_\mu \right] S(p) \right): \,.
\label{HQE_6}
\end{eqnarray}

\noindent
Subsequently we take the limit $y \rightarrow 0$, i.e. no time-ordering anymore, and obtain the important relation 
\begin{eqnarray}
\langle \Omega | \bar{q} \, {\cal O}\left[\overrightarrow{D}_\mu\right] \, q | \Omega \rangle &=& 
\langle \Omega | : \bar{q}\, {\cal O}\left[\overrightarrow{D}_\mu\right] q: | \Omega \rangle
\nonumber\\
&& - i \int \frac{d^4p}{(2\pi)^4}
\langle \Omega | : \TrCD \left({\cal O} \left[-ip_\mu - i \tilde{A}_\mu\right] S(p)\right) : | \Omega \rangle \,,
\label{definition_physical_condensates}
\end{eqnarray}

\noindent
where we have also taken the expectation value. So far, the relation 
(\ref{definition_physical_condensates}) has been applied to the in-medium case 
in \cite{Zschocke2,Hilger_Diplom,D_Meson1,D_Meson2,Hilger_Kampfer_Leupold,HilgerA,HilgerB}. Here, 
${\cal O}\left[-ip_\mu - i \tilde{A}_\mu\right]$ denotes the Fourier transformed operator function, where $\tilde{A}_\mu$ 
is defined by Eq.~(\ref{eq:def_a_tilde}). The derivatives are now contained in the gluon fields. Therefore, the ordering 
among the Lorentz indices is important as the field operators act on everything to the right. 
We define (\ref{definition_physical_condensates}) as the relation between physical condensates and normal-ordered 
condensates. Renormalizing the physical condensates by absorbing the divergent terms of $\Pi^{(0)}_{G^2}$ on the r.h.s. of 
Eq.~(\ref{definition_physical_condensates}) cancels out the infrared divergences and yields infrared stable Wilson 
coefficients. We note that the extension of operator mixing for the four-quark condensates can straightforward be obtained 
by very similar steps like (\ref{HQE_5}) - (\ref{definition_physical_condensates}).

Eq.~(\ref{definition_physical_condensates}) is valid to any loop-order of the corresponding Feynman diagrams. Here, for our 
purposes it will be sufficient to calculate the matrix elements only up to one-loop order, denoted by the index $^{(1)}$. 
Up to order ${\cal O} \left(\alpha_s\right)$ the following set of relations has been obtained in $\overline{\rm MS}$ scheme: 
\begin{eqnarray}
\langle \Omega | \bar{q} q | \Omega \rangle^{(1)} &=& \langle \Omega | : \bar{q} q : | \Omega \rangle ^{(0)}
+ \frac{3}{4\pi^2}m_q^3 \left( \ln{\frac{\mu^2}{m_q^2}}+1 \right) 
\nonumber\\
&& - \frac{1}{12m_q} \langle \Omega | : \frac{\alpha_s}{\pi}\;G^2 : |\Omega\rangle^{(0)}\,,
\label{physical_condensates_2a}
\\
\nonumber\\
\langle \Omega | \bar{q}\,g_s\,\sigma\,{\cal G}\,q | \Omega \rangle^{(1)} &=& 
\langle \Omega | : \bar{q}\,g_s\,\sigma\,{\cal G}\,q : | \Omega \rangle^{(0)}
- \frac{1}{2}m_q \ln \frac{\mu^2}{m_q^2} \,\langle \Omega | : \frac{\alpha_s}{\pi}\,G^2 : |\Omega\rangle^{(0)} ,
\nonumber\\
\label{physical_condensates_2b}
\\ 
\nonumber\\
\langle \Omega | \bar{q}\gamma_{\mu} \,i\,\overrightarrow{D}_{\nu}\, q | \Omega \rangle^{(1)} &=& 
\langle \Omega | : \bar{q}\gamma_{\mu} \,i \,\overrightarrow{D}_{\nu}\,q: | \Omega \rangle^{(0)}
+ \frac{3}{16\pi^2}m_q^4 g_{\mu\nu} \left( \ln{\frac{\mu^2}{m_q^2}} + 1 \right) 
\nonumber\\
&& \hspace{-3.0cm} - \frac{g_{\mu\nu}}{48} \; \langle \Omega | : \frac{\alpha_s}{\pi}\;G^2 : |\Omega\rangle^{(0)}
\nonumber\\
&& \hspace{-3.0cm} 
+ \frac{1}{18} \left( g_{\mu\nu} - 4\frac{v_\mu v_\nu}{v^2} \right) \left( \ln{\frac{\mu^2}{m_q^2}} - \frac{1}{3} \right) 
\langle \Omega | : \frac{\alpha_s}{\pi}\left(\frac{\left(v G\right)^2}{v^2} - \frac{G^2}{4}\right) : |\Omega\rangle^{(0)}
\,,
\label{physical_condensates_2c}
\\ 
\nonumber\\
\langle \Omega | \bar{q} \,i \,\overrightarrow{D}_{\mu}\, i\, \overrightarrow{D}_{\nu}\, q | \Omega \rangle^{(1)} 
&=& \langle \Omega | : \bar{q} \,i \,\overrightarrow{D}_{\mu} \,i \,\overrightarrow{D}_{\nu} \, q : | \Omega \rangle^{(0)}
+ \frac{3m_q^5}{16\pi^2} g_{\mu\nu} \left( \ln{\frac{\mu^2}{m_q^2}} + 1 \right)
\nonumber\\
&& \hspace{-3.0cm} 
+ \frac{m_q}{16}g_{\mu\nu} \left( \ln{\frac{\mu^2}{m_q^2}} - \frac{1}{3} \right) \langle \Omega | : \frac{\alpha_s}{\pi}\;G^2 : |\Omega\rangle^{(0)}
\nonumber\\
&& \hspace{-3.0cm} 
- \frac{m_q}{36} \left( g_{\mu\nu} - 4\frac{v_\mu v_\nu}{v^2} \right) \left( \ln{\frac{\mu^2}{m_q^2}} + \frac{2}{3} \right) 
\langle \Omega | : \frac{\alpha_s}{\pi}\left(\frac{\left(v G\right)^2}{v^2} - \frac{G^2}{4}\right) : |\Omega\rangle^{(0)}\,.
\label{physical_condensates_2d}
\end{eqnarray}

\noindent
In Appendix \ref{Appendix_Mixing} relation (\ref{physical_condensates_2d}) is proven in some detail. 
Furthermore, we notice the relations 
\begin{eqnarray}
 \langle \Omega | \frac{\alpha_s}{\pi}\;G^2  | \Omega \rangle^{(1)} &=&
\langle \Omega | : \frac{\alpha_s}{\pi}\;G^2 : | \Omega \rangle^{(0)}\,,
\label{Gluon_Operator_1}
\\
\nonumber\\
\langle\Omega|\frac{\alpha_s}{\pi}\left(\frac{\left(v\;G\right)^2}{v^2} - \frac{G^2}{4}\right)|\Omega \rangle^{(1)}
&=& 
\langle\Omega |:\frac{\alpha_s}{\pi}\;\left(\frac{\left(v\;G\right)^2}{v^2} - \frac{G^2}{4}\right):|\Omega\rangle^{(0)}\,,
\label{Gluon_Operator_2}
\end{eqnarray}

\noindent
because these operators are already of order ${\cal O} \left(\alpha_s\right)$. 
The self-consistency of the relations (\ref{physical_condensates_2c}) and (\ref{physical_condensates_2d}) with the 
relations (\ref{physical_condensates_2a}) and (\ref{physical_condensates_2b}) can easily 
be verified by a contraction with the metric tensor and with the aid of the equation of motion 
(\ref{EOM1}) and relation (\ref{Relation_B}), respectively. 

The relations for the scalar operators 
(\ref{physical_condensates_2a}) and (\ref{physical_condensates_2b}) are identical to the relations for the vacuum 
condensates and were given, for instance, in \cite{Jamin_Munz,Broadhurst1,Narison2}; 
in vacuum higher orders of (\ref{physical_condensates_2a}) were obtained in \cite{Broadhurst1}.
For the other condensates which vanish in vacuum we obtain of course 
$\langle \overline{q} \gamma_{\mu} q \rangle^{(1)} = \langle : \overline{q} \gamma_{\mu} q : \rangle^{(0)}$, 
$\langle \overline{q} \,i\,\overrightarrow{D}_{\mu} q \rangle^{(1)} = \langle : \overline{q} 
\,i\,\overrightarrow{D}_{\mu} q : \rangle^{(0)}$, and  
$\langle \bar{q}\,\gamma_5\,\gamma_{\alpha}\,G_{\mu\nu}\,q \rangle^{(1)} =
\langle : \bar{q}\,\gamma_5\,\gamma_{\alpha}\,G_{\mu\nu}\,q : \rangle^{(0)}$. 

From now on we drop the label $^{(1)}$, i.e. non-normal-ordered condensates are always on one-loop-level throughout 
the work. We note, that a scale $\mu$ naturally appears when the non-normal-ordered condensates are introduced, and they 
would even not be well-defined without $\mu$. The $\mu$ scale is an IR cutoff which 
marks the region between perturbative and non-perturbative physics. 
Hence, any sensitivity to the light-quark mass must be accounted for by a 
redefinition of the condensates. This resembles the case of parton distribution 
functions (PDFs) in deep inelastic scattering. The next-to-leading-order calculations 
are IR divergent which can be accounted for by a redefined PDF. As a consequence
the PDFs become scale dependent which gives rise to the famous DGLAP
(Dokshitzer, Gribov, Lipatov, Altarelli, Parisi) equations \cite{DGLAP1,DGLAP2,DGLAP3}. We underline that a careful 
application of these operator mixing relations (\ref{physical_condensates_2a}) - (\ref{Gluon_Operator_2}) is 
mandatory for a consistent treatment of the OPE if one internal quark mass of the currents is much smaller than the 
confinement scale $\Lambda_{\rm QCD}$.

\subsection{Insertion of operator mixing}

By inserting Eqs.~(\ref{physical_condensates_2a}) - (\ref{Gluon_Operator_2}) into 
Eq.~(\ref{before_Lorentz_projection_d}) we obtain $\Pi_d^{(2)}(q)$ in terms of non-normal-ordered condensates:
\begin{eqnarray}
\Pi_d^{(2)}(q) &=& 
\langle \Omega | \bar{d}d | \Omega \rangle \frac{m_c}{q^2-m_c^2}
- \langle \Omega | \bar{d} \, i \overrightarrow{D}_{\mu} d | \Omega \rangle \frac{2 m_c q^{\mu}}{(q^2-m_c^2)^2}
\nonumber\\
&& \hspace{-1.5cm} - \langle \Omega | \bar{d} \gamma_{\lambda} d | \Omega \rangle \frac{q^{\lambda}}{q^2-m_c^2}
+ \langle \Omega | \bar{d} \gamma_{\lambda} i \overrightarrow{D}_{\mu} d | \Omega \rangle
2 \frac{q^{\mu} q^{\lambda}}{(q^2-m_c^2)^2}
\nonumber\\
&& \hspace{-1.5cm} - \langle \Omega | \bar{d} \, \overrightarrow{D}_{\mu} \overrightarrow{D}_{\nu} d | \Omega \rangle
4 \frac{m_c q^{\mu} q^{\nu}}{(q^2-m_c^2)^3} 
+ \langle \Omega | \bar{d} \gamma_{\lambda} \overrightarrow{D}_{\mu} \overrightarrow{D}_{\nu} d | \Omega \rangle
4 \frac{q^{\mu} q^{\nu} q^{\lambda}}{(q^2-m_c^2)^3}
\nonumber\\
&& \hspace{-1.5cm} - \langle \Omega | \bar{d} \,g_s\, \gamma_{\alpha}\,\sigma\,{\cal G}\,d | \Omega \rangle
\frac{1}{2} \frac{ q_\alpha}{(q^2-m_c^2)^2}
\nonumber\\
&& \hspace{-1.5cm} + \GGVren \left( \frac{1}{12} \frac{m_c}{m_d} \frac{1}{q^2-m_c^2} 
+ \frac{1}{24} \frac{m_c^2}{(q^2-m_c^2)^2} - \frac{1}{24} \frac{1}{q^2-m_c^2} \right)
\nonumber\\
&& \hspace{-1.5cm} - \GGMren
\frac{1}{9}\;
\frac{1}{(q^2-m_c^2)^2} \left( q^2 - 4 \frac{(vq)^2}{v^2} \right) \left( \ln \frac{\mu^2}{m_d^2} - \frac{1}{3} \right).
\label{quark_condensate_with_physical_OPE}
\end{eqnarray}

\noindent
For the mixed quark-gluon condensate (next to last term in Eq.~(\ref{before_Lorentz_projection_d})) we have used relation 
(\ref{Relation_B}), while for the mixed quark-gluon condensate which contains $\gamma_5$ (last term in 
Eq.~(\ref{before_Lorentz_projection_d})) we have used the relation (\ref{arbitrary_fourvector}) which can be shown to be 
valid for an arbitrary four-vector. Special care is needed when applying the equation of motion (\ref{EOM1}) to some 
condensates in (\ref{before_Lorentz_projection_d}) and neglecting terms $\propto m_d$, because the operator mixing may 
introduce terms which cancel out factors of $m_d$. For instance, this is the case for 
$\langle : \bar{d} \gamma_{\lambda} \overrightarrow{D}_{\mu} d : \rangle$; contraction with the metric tensor and 
application of equation of motion (\ref{EOM1}) results in a term $\propto m_d \langle : \bar{d}d : \rangle$, which, however, 
remains finite and non-zero after operator mixing. Thus, we cannot neglect such 
terms.

Here, we emphasize that the both IR-divergent terms $\propto m_d^{-1}$ and 
$\propto \ln m_d$ in Eq.~(\ref{quark_condensate_with_physical_OPE}) will cancel the 
corresponding IR-divergent terms in (\ref{Pi_G2}). This cancellation will be the topic of the next Section. 

\section{OPE in the limit of a vanishing light-quark mass \label{sct:Limit}}

A sensitivity to light-quark mass signals a sensitivity to very small quark and gluon momenta.
Physically such momenta are cut off by confinement, i.e. at a scale of 
$\Lambda_{\rm QCD} \simeq 245\,{\rm MeV}$, and not by
the light current quark mass at a few ${\rm MeV}$ scale. Hence, any sensitivity to the quark mass must be accounted
for by a redefinition of the condensates. Moreover, the limit $m_d \rightarrow 0$ is important, since it demonstrates 
the cancellation of all IR-divergences in (\ref{Pi_G2}) and (\ref{quark_condensate_with_physical_OPE}), which is 
a fundamental feature for the consistency of the whole approach.

In order to calculate the complete OPE for the D-meson in medium with infrared stable Wilson coefficients in terms of 
non-normal-ordered condensates we have to add the perturbative part (\ref{Im_Pert_5}), the expressions (\ref{Pi_G2}) by  
virtue of Eqs.~(\ref{Gluon_Operator_1}) and (\ref{Gluon_Operator_2}) and the terms 
in Eq.~(\ref{quark_condensate_with_physical_OPE}).  
We emphasize the cancellation of the IR-divergent term $\propto \ln(m_d^2)$ in the Wilson coefficient of the 
medium-specific gluon condensate and the IR-divergent term $\propto m_d^{-1}$ in the Wilson coefficient of the gluon 
condensate. This important result of cancellation of IR-divergences due operator mixing 
(\ref{definition_physical_condensates}) has been demonstrated at the first time in \cite{Zschocke2} 
for in-medium D-mesons at rest. In a more comprehensive analysis this result has been extended to the case of 
in-medium D-mesons with finite three-momenta in \cite{Hilger_Diplom}. 

Now we will take the limit $m_d \rightarrow 0$ 
which shows explicitly that the complete OPE in terms of non-normal-ordered condensates is infrared stable. 
The complete projections with respect to Dirac- and color-indices of the condensates onto invariant structures
are given in Appendix \ref{projections}. By means of these projections, we obtain up to mass dimension 5:
\begin{eqnarray}
\Pi (q) &=& \Pi^{\rm pert}(q) + \langle \Omega | \bar{d}d | \Omega \rangle \frac{m_c}{q^2-m_c^2}
- \langle \Omega | \bar{d}g\sigma{\cal G}d | \Omega \rangle
\frac{1}{2}\,\frac{m_c\,q^2}{(q^2-m_c^2)^3}
\nonumber\\
&& \hspace{-1.0cm}
- \langle \Omega | \frac{\alpha_s}{\pi}\,G^2 | \Omega \rangle  \frac{1}{12} \frac{1}{q^2-m_c^2}
\nonumber\\
&& \hspace{-1.0cm} 
+ \langle \Omega | \frac{\alpha_s}{\pi} \left(\frac{\left(v \, G\right)^2}{v^2} - \frac{G^2}{4}\right) | \Omega \rangle
\left( q^2 - 4\frac{(vq)^2}{v^2} \right)
\nonumber\\
&& \hspace{-1.0cm} \times \Bigg( - \frac{7}{54}\,\frac{1}{(q^2-m_c^2)^2} - \frac{1}{9}\,\frac{1}{(q^2-m_c^2)^2}\,
\ln \frac{\mu^2}{m_c^2} - \frac{2}{9} \,\frac{1}{(q^2-m_c^2)^2}\,
\ln \left( - \frac{m_c^2}{q^2-m_c^2} \right)\Bigg)
\nonumber\\
&& \hspace{-1.0cm} - \langle \Omega | \bar{d} \hat{v} \frac{(vi\overrightarrow{D})}{v^2} d | \Omega \rangle
\left( q^2 - 4\frac{(vq)^2}{v^2} \right) \frac{2}{3}\,\frac{1}{(q^2-m_c^2)^2}\,
\nonumber\\
&& \hspace{-1.0cm} - \langle \Omega | \bar{d} \hat{v} (v \overrightarrow{D})^2 d | \Omega \rangle
\frac{vq}{v^4} \left( q^2 - 2\frac{(vq)^2}{v^2} \right) \,4\,\frac{1}{(q^2-m_c^2)^3}
\nonumber\\
&& \hspace{-1.0cm} + \langle \Omega | \bar{d} \hat{v} g_s \sigma {\cal G} d | \Omega \rangle 
\frac{vq}{v^2} \left( \frac{2}{3} \frac{q^2 - \frac{(vq)^2}{v^2}}{(q^2-m_c^2)^3}
- \frac{1}{(q^2-m_c^2)^2} \right) - \langle \Omega | \bar{d} \hat{v} d | \Omega \rangle \frac{vq}{v^2} \frac{1}{q^2-m_c^2}
\nonumber\\
&& \hspace{-1.0cm} + \left[ \frac{1}{3} \langle \Omega | \bar{d} \frac{(v\overrightarrow{D})^2}{v^2} d | \Omega \rangle 
- \frac{1}{24} \langle \Omega | \bar{d} g_s \sigma {\cal G} d | \Omega \rangle \right]
\left( q^2 - 4\frac{(vq)^2}{v^2} \right) \,4\,\frac{m_c}{(q^2-m_c^2)^3}\,.
\label{OPE}
\end{eqnarray}

\noindent
The angled brackets in the last term indicate that this combination vanishes in vacuum, thus 
denotes a medium-specific part: applying vacuum projections to this medium specific term makes it zero in the vacuum limit. 
Numerical values of the condensates in (\ref{OPE}) are given in \cite{D_Meson1}. 
The IR-stable OPE in Eq.~(\ref{OPE}) is the main result of our investigation. It is valid for a meson whith the 
four-momentum $q^{\mu} = (q_0, \ve{q})$, while the surrounding medium has a four-velocity $v^{\mu} = (1, \ve{v})$ 
in respect to a given frame.

Needless to say, in the limit $m_d \rightarrow 0$ the OPE simplifies considerably. But we emphasize that the limit 
$m_d \rightarrow 0$ is not performed because such terms are small anyway or in order to simplify the OPE, instead this 
limit is necessary from physical reasons: the dependence of the correlator from the light-quark mass $m_d$ 
signals a sensitivity to very small momenta, which are cut off by confinement, that means at a scale 
of $\Lambda_{\rm QCD}$, which is much larger than the light-quark mass. Thus only the OPE in the limit of a vanishing 
light-quark mass is meaningful from the physical point of view.

\section{OPE for a meson at rest with respect to the medium \label{sct:rest_frame}}

In the previous Section we have presented the OPE for the general case $q^{\mu} = (q_0,\ve{q})$. However, for many 
investigations concerning in-medium properties of D-mesons it will be sufficient to consider the D-mesons at rest with 
respect to the medium they are embedded in. Therefore, we will also consider the special case 
$q^{\mu} = (q_0, \ve{0})$; furthermore we choose a frame comoving with the medium $v^{\mu} = (1,\ve{0})$.  
We present the OPE separated into an even part and an odd part: 
$\Pi(q_0) = \Pi^{\rm even} (q_0^2) + q_0\,\Pi^{\rm odd} (q_0^2)$. Furthermore, we perform an 
analytical continuation $q_0 = i \, \omega$. 
Then, the even part of the IR-stable OPE for $m_d \rightarrow 0$ can be written in the form:
\begin{eqnarray}
\Pi^{\rm even} (\omega) &=& C_0 (\omega) - \langle \Omega | \bar{d}d | \Omega \rangle \,\frac{m_c}{\omega^2 + m_c^2} 
- \langle \Omega | \bar{d}\,g_s\,\sigma{\cal G}\,d | \Omega \rangle \frac{1}{2}\, \frac{m_c\,\omega^2}{(\omega^2+m_c^2)^3} 
\nonumber\\
&& \hspace{-2.0cm} + \frac{1}{12}\,\langle \Omega | \frac{\alpha_s}{\pi}\,G^2 | \Omega \rangle\,\frac{1}{\omega^2 + m_c^2}  
\nonumber\\
&& \hspace{-2.0cm} - \langle \Omega | \frac{\alpha_s}{\pi} \left( \frac{\left(v G \right)^2}{v^2} - \frac{G^2}{4}\right) | \Omega \rangle 
\left( \frac{7}{18} + \frac{1}{3} \,\ln \frac{\mu^2}{m_c^2} + \frac{2}{3}\,\ln \left(\frac{m_c^2}{\omega^2 + m_c^2}\right) \right) \frac{\omega^2}{\left(\omega^2 + m_c^2\right)^2} 
\nonumber\\
&& \hspace{-2.0cm} - 2 \langle \Omega | d^{\dagger} i \overrightarrow{D}_0 d | \Omega \rangle\, 
\frac{\omega^2}{\left(\omega^2 + m_c^2\right)^2} 
- 4 \left[  \langle \Omega | \overline{d} \overrightarrow{D}_0^2 d | \Omega \rangle  
-  \langle \Omega | \frac{1}{8} \overline{d} g_s \sigma {\cal G} d | \Omega \rangle \right] 
\frac{m_c \omega^2}{\left(\omega^2 + m_c^2\right)^3} \,.
\nonumber\\
\label{vanishing_momentum_even}
\end{eqnarray}
  
\noindent
Note, the angled brackets in the last term denote that this combination vanishes in vacuum, thus denotes a 
medium-specific part which is absent in vacuum. 
The odd part of the IR-stable OPE for $m_d \rightarrow 0$ is given by:
\begin{eqnarray}
 \Pi^{\rm odd} (\omega) &=& \langle \Omega | d^{\dagger} d | \Omega \rangle\, \frac{1}{\omega^2 + m_c^2} 
+ 4\,\langle \Omega | d^{\dagger}\,\overrightarrow{D}_0^2\, d | \Omega \rangle\,
\frac{\omega^2}{\left(\omega^2 + m_c^2\right)^3} 
\nonumber\\
&& - \langle \Omega | d^{\dagger} g_s\,\sigma\,{\cal G}\,d | \Omega \rangle\,
\frac{1}{\left(\omega^2 + m_c^2\right)^2}\,.
\label{vanishing_momentum_odd}
\end{eqnarray}

\noindent
The OPE is an asymptotic series. As such it must be truncated at a certain 
mass dimension and, therefore, takes into account a finite number of operators. One way to deal with asymptotic series is 
to perform a Borel transformation which suppresses the effect of higher mass dimensional operators. The Borel transformation of a function $f (Q^2)$ is defined by 
\begin{eqnarray}
 {\cal B} \left[ f \left(M^2\right) \right] &=& \lim_{n \rightarrow \infty}\; \lim_{Q^2 \rightarrow n\,M^2} \;
\frac{\left(Q^2\right)^{n+1}}{n!} \left( - \frac{d}{d Q^2}\right)^n\; f \left(Q^2\right),
\label{Borel_Transformation}
\end{eqnarray}

\noindent
where the parameter $M$ is the so-called Borel mass. Applying the Borel transformation to the even part 
(\ref{vanishing_momentum_even}) we obtain 
\begin{eqnarray}
{\cal B} \left[\Pi^{\rm even} (\omega^2) \right] &=& 
\frac{1}{\pi} \int\limits_{m_c^2}^{\infty} ds\;{\rm e}^{ - s/M^2}\;{\rm Im} \Pi^{\rm pert} (s) 
\nonumber\\
&& \hspace{-3.0cm}
+ {\rm e}^{ - m_c^2/M^2}\;\Bigg(-m_c \langle \Omega | \overline{d} d | \Omega \rangle + \frac{1}{2} 
\left(\frac{m_c^3}{2\,M^4} - \frac{m_c}{M^2}\right)
\langle \Omega | \overline{d} \,g_s\,\sigma\,{\cal G}\,d | \Omega \rangle  + \frac{1}{12}\,\langle \Omega | \frac{\alpha_s}{\pi}\,G^2 | \Omega \rangle
\nonumber\\
&& \hspace{-3.0cm}
+ \left[\left(\frac{7}{18} + \frac{1}{3}\,\ln \frac{\mu^2\,m_c^2}{M^4} - \frac{2}{3}\,\gamma_E\right)\left( \frac{m_c^2}{M^2}- 1\right) 
- \frac{2}{3}\,\frac{m_c^2}{M^2}\right] \,
\langle \Omega | \frac{\alpha_s}{\pi} \left( \frac{\left(v G \right)^2}{v^2} - \frac{G^2}{4}\right) | \Omega \rangle
\nonumber\\
&&\hspace{-3.0cm} 
+ 2 \left(\frac{m_c^2}{M^2} - 1\right) \langle \Omega | d^{\dagger}\,i\,\overrightarrow{D}_0\, d | \Omega \rangle
\nonumber\\
&& \hspace{-3.0cm}  + 4\,\left(\frac{m_c^3}{2\,M^4} - \frac{m_c}{M^2}\right)
 \left[ \langle \Omega | \overline{d}\,\overrightarrow{D}_0^2\,d | \Omega \rangle  
-  \langle \Omega | \frac{1}{8}\,\overline{d}\,g_s \,\sigma\,{\cal G}\,d | \Omega \rangle\right]\Bigg),
\label{Borel_Transformation_even}
\end{eqnarray}

\noindent
while for the odd part (\ref{vanishing_momentum_odd}) we obtain 
\begin{eqnarray}
{\cal B} \left[\Pi^{\rm odd} (\omega^2) \right] &=&  {\rm e}^{- m_c^2/M^2}\;\Bigg( 
\langle \Omega | d^{\dagger} d | \Omega \rangle - 4 \left(\frac{m_c^2}{2\,M^4} - \frac{1}{M^2}\right) 
\langle \Omega | d^{\dagger}\,\overrightarrow{D}_0^2\, d | \Omega \rangle
\nonumber\\
&& - \frac{1}{M^2}\,\langle \Omega | d^{\dagger}\,g_s\,\sigma\,{\cal G}\, d | \Omega \rangle\Bigg).
\label{Borel_Transformation_odd}
\end{eqnarray}

\noindent
This Borel transformed OPE has been presented in \cite{D_Meson1,D_Meson2} for D-mesons in-medium, however no further 
details about how to arrive at this correct OPE have been presented; so far a detailed presentation can only be found 
in \cite{Zschocke2} and in the more extensive study \cite{Hilger_Diplom}, where the approach of operator mixing and 
cancellation of IR-divergences for the vacuum case \cite{Chetyrkin2,Jamin_Munz,Broadhurst1} has been generalized to the 
in-medium case. 

\section{Comparison with the literature} \label{sct:Literature}

Even the OPE for D-mesons in vacuum is not as trivial as the common belief might be. This can be illustrated, 
for instance, by a brief review of the history about the different results obtained for the Wilson coefficient of the 
condensate $\langle {\rm vac} | g_s\,\overline{d}\,\sigma\,{\cal G}\,d | {\rm vac} \rangle$. The first attempt to 
calculate this coefficient was done in \cite{Vacuum_1}, where a wrong factor $- 1/4$ has been presented. 
At the first time, the correct result $- 1/2$ has been presented in \cite{Aliev}, 
where the needed correction of \cite{Vacuum_1} was explicitly mentioned. 
Later, in \cite{Vacuum_2} a wrong factor 
$+ 1/4$ has been given. The correct result in Ref. \cite{Aliev} has later 
been confirmed in \cite{Jamin_Munz,Hayashigaki2}. But this was not the end of the story. In \cite{Vacuum_3} a wrong 
factor has been given again, which has later been corrected in \cite{Vacuum_4} by the same author, explicitly mentioning the 
needed corrections in Refs. \cite{Vacuum_2,Vacuum_3}. This brief survey of history shows, that the calculation of 
the OPE for D-mesons needs special care, and even more for the in-medium case. 

So far, in-medium QSR for D-mesons were given in \cite{Morath1,Morath2,Hayashigaki1} and in our own 
investigations \cite{Zschocke2,Hilger_Diplom,D_Meson1,D_Meson2}. The references \cite{Morath1} and \cite{Morath2} are 
from the same author(s) and the given results agree with each other. Furthermore, in a recent study \cite{Hayashigaki2} the 
D-mesons in vacuum were considered, but since the Wilson coefficients of scalar condensates in vacuum and in-medium are 
the same we can compare the results. Thus, at the moment being a comparison of our results is meaningful with the findings 
of Refs.~\cite{Hayashigaki1,Morath1,Hayashigaki2}. 

\subsection{Scalar part of the OPE for D-mesons in medium}

In Table \ref{table1} we compare our results for scalar condensates in-medium with the corresponding results of 
Refs.~\cite{Hayashigaki1,Morath1,Hayashigaki2}. 
Here, we list the Borel transformed Wilson coefficients $c_{\cal O} / {\rm e}^{-m_c^2/M^2}$. 
Obviously, there are serious disagreements of the findings in \cite{Hayashigaki1,Hayashigaki2} 
to our results, while the results of Ref.~\cite{Morath1} do agree. It seemed to us, that the wrong Wilson coefficient 
for the scalar gluon condensate in \cite{Hayashigaki1,Hayashigaki2} is either caused by a missing sign in an intermediate 
step somewhere in their calculation or because of an incorrect operator mixing in Refs.~\cite{Hayashigaki1,Hayashigaki2}.

\begin{table}[!h]
\begin{tabular}{| c | c | c | c |}
\hline
&&&\\[-20pt]
&\hbox to 20mm{\hfill $ \langle \Omega | \overline{d} d | \Omega \rangle $ \hfill}
&\hbox to 20mm{\hfill $\langle  \Omega | \frac{\alpha_s}{\pi} G^2  | \Omega \rangle $\hfill}
&\hbox to 20mm{\hfill $\langle  \Omega | \overline{d} g_s \sigma {\cal G} d  | \Omega \rangle $ \hfill}\\[3pt]
\hline
&&&\\[-20pt]
Eq.~(\ref{Borel_Transformation_even}) & $- m_c$ & $\frac{1}{12}$ 
& $ - \frac{1}{2}\,\frac{1}{M^2} \left( 1- \frac{1}{2}\,\frac{m_c^2}{M^2} \right) m_c $  \\[3pt]
\hline
Ref.~\cite{Hayashigaki1} & $ - m_c$ & $\frac{1}{12} - \frac{1}{24}\,\frac{m_c^2}{M^2}$
&  \\[3pt]
\hline
Ref.~\cite{Morath1} & $- m_c$  & $\frac{1}{12}$
& $ - \frac{1}{2}\,\frac{1}{M^2} \left( 1- \frac{1}{2}\,\frac{m_c^2}{M^2} \right) m_c $ \\[3pt]
\hline
Ref.~\cite{Hayashigaki2} & $- m_c$ & $\frac{1}{12}\,\left(\frac{3}{2} - \frac{m_c^2}{M^2} \right) $ 
& $ - \frac{1}{2}\,\frac{1}{M^2} \left( 1- \frac{1}{2}\,\frac{m_c^2}{M^2} \right) m_c $ \\[3pt]
\hline
\end{tabular}
\caption{Comparison of scalar condensates of the OPE given by Eq.~(\ref{Borel_Transformation_even}) with 
Refs.~\cite{Hayashigaki1,Morath1,Hayashigaki2}. In \cite{Hayashigaki1} there was no scalar mixed condensate, 
because only operators up to mass dimension 4 were taken into account.}
\label{table1}
\end{table}

\subsection{Tensor part of the OPE for D-mesons in medium}

In Table \ref{table2}, we compare the tensor part of the in-medium OPE with the literature. As before, we list the Borel 
transformed Wilson coefficients $c_{\cal O} / {\rm e}^{-m_c^2/M^2}$. We have found a serious disagreement between our 
results and the findings of Ref.~\cite{Hayashigaki1}. Unfortunately, it is impossible to compare our results directly 
with the ones given in \cite{Morath1,Morath2}, as it is somehow hidden in their work. We could not understand in detail 
how the authors of \cite{Morath1,Morath2} have treated the IR singularities. However, for the unprojected 
OPE and before introducing physical condensates such a comparison is possible. That means we compare our result 
(\ref{before_Lorentz_projection_d}) with \cite{Morath1}, although this is less meaningful, because 
many calculations still have to be done from this point on. We have found 
an agreement of (\ref{before_Lorentz_projection_d}) with \cite{Morath1} to a large extent, except a slight disagreement of a 
factor $\frac{1}{3}$ in the Wilson coefficient of the condensate 
$\langle : \overline{d} \gamma_{\lambda} \sigma \,{\cal G} d : \rangle$, which emerges from
$\langle : \bar{d} \, \overrightarrow{D}_{\mu} \overrightarrow{D}_{\nu} d : \rangle$ and 
$\langle : \bar{d} \gamma_5 \gamma_\alpha {\cal G}_{\mu\nu} d : \rangle$ by means of the equation of motion 
(\ref{EOM1}); for further details see \cite{Hilger_Diplom}. 

\begin{table}[!h]
\begin{tabular}{| c | c | c | }
\hline
&&\\[-20pt]
&$\langle  \Omega | \overline{d}^{\dagger}\,i\,\overrightarrow{D}_0\, d  | \Omega \rangle $ & 
$\langle \Omega | \frac{\alpha_s}{\pi} \left( \frac{\left(v G \right)^2}{v^2} - \frac{G^2}{4}\right)  | \Omega \rangle$ 
\\[3pt]
\hline
&&\\[-20pt]
Eq.~(\ref{Borel_Transformation_even}) 
& $2 \left( \frac{m_c^2}{M^2} - 1 \right)$ & $\left(\frac{7}{18} + \frac{1}{3}\,\ln \frac{\mu^2\,m_c^2}{M^4} - \frac{2}{3}\,\gamma_E \right) \left( \frac{m_c^2}{M^2} - 1\right)- \frac{2}{3}\,\frac{m_c^2}{M^2}$ \\[3pt]
\hline
Ref.~\cite{Hayashigaki1} 
& $2 \left( \frac{m_c^2}{M^2} - 1 \right)$ & $\frac{1}{3} \bigg[ \frac{4}{3} - \frac{1}{6}\,\frac{m_c^2}{M^2} + \frac{1}{2}\,\frac{m_c^6}{M^6} + \left(1 - \frac{m_c^2}{M^2}\right) \ln
\left(\frac{m_c^2}{4\,\pi\,\mu^2}\right)$ \\[3pt]
&& $+ {\rm e}^{m_c^2/M^2} \left(-2\,\gamma_E - \ln \frac{m_c^2}{M^2} + \int\limits_{0}^{m_c^2/M^2} dt\,\frac{1 - {\rm e}^{-t}}{t}\right)
\bigg]$ \\[3pt]
\hline
\end{tabular}
\caption{Comparison of tensorial condensates of OPE given by Eq.~(\ref{Borel_Transformation_even}) with 
Ref.~\cite{Hayashigaki1}.}
\label{table2}
\end{table}

\noindent
Finally, we note that for the odd part of the in-medium OPE a comparison of our result with \cite{Hayashigaki1} 
is also not possible, because it is not clear whether the odd part has been considered 
in \cite{Hayashigaki1} or not; note there is no odd part in vacuum, hence a comparison with \cite{Hayashigaki2} 
is impossible. Furthermore, in \cite{Morath1,Morath2} the OPE is given in terms of normal-ordered condensates 
which implies IR-divergent Wilson coefficients and makes a comparison difficult. 

In summary, we come to 
the conclusion that the OPE for in-medium D-mesons has not been determined correctly, except by the presentations 
given in \cite{D_Meson1,D_Meson2} where, however, no further details of the involved calculations have been given; 
so far, such details were only presented in \cite{Zschocke2} and in the more detailed analysis \cite{Hilger_Diplom}. 

%================================================
\section{Summary}\label{sct:summary}
%================================================ 

We have determined the in-medium OPE, at first non-trivial order in powers of the QCD coupling constant 
and including all operators up to mass dimension 5, for heavy-light-quark pseudoscalar mesons: 
D-mesons and B-mesons. We have outlined that it is sufficient to consider the OPE
just for $D^{+}$ mesons, since all other OPE's of heavy-light-quark 
pseudoscalar mesons can easily be deduced from that result.

So far, in-medium QSR for D-mesons are fairly rare 
\cite{Morath1,Morath2,Hayashigaki1,Zschocke2,Hilger_Diplom,D_Meson1,D_Meson2,Hilger_Kampfer_Leupold,HilgerA},  
and the OPE in these works differ significantly. Especially, the applied OPE in \cite{Morath1,Morath2,Hayashigaki1} is 
incorrect. From our view, the reason for this fact is that the derivation of the correct OPE turns out to be 
an ambitious assignment of a task. Accordingly, it is timely to present the derivation of the OPE for 
D-mesons in matter in some detail.  

First, we have determined the OPE with the aid of Wick's theorem, leading to an OPE 
in terms of normal-ordered condensates, see Eq.~(\ref{OPE_IR_unstable}). 
The complete OPE up to operators of mass dimension 5 
in terms of normal-ordered condensates is given by Eq.~(\ref{eq:Wick_Theorem_0}) and Eq.~\ref{Decomposition}: 
the term $\Pi^{\rm pert}$ by Eqs.~(\ref{Im_Pert_5}) and (\ref{Im_Pert_10}), the term $\Pi_{G^2}^{(0)}$ 
by Eq.~(\ref{Pi_G2}) and the term $\Pi_d^{(2)}$ by 
Eq.~(\ref{before_Lorentz_projection_d}). We also have determined 
$\Pi_c^{(2)}$ given by Eq.~(\ref{before_Lorentz_projection_c}) which contains 
charmed condensates, but we have argued why these charmed condensates can be neglected. 
The Wilson coefficients of OPE (\ref{OPE_IR_unstable}) are IR-divergent, that means they are infinite in the limit 
of a vanishing light-quark mass, $m_d \rightarrow 0$. 

It has been described in detail that a consistent treatment of the OPE is obtained in terms of non-normal-ordered 
condensates, see Eq.~(\ref{OPE_IR_stable}). The relation between normal-ordered condensates and non-normal-ordered 
condensates is given by Eq.~(\ref{definition_physical_condensates}) to any loop-order. We have determined explicitly 
the relations among these condensates to one-loop-order, that is nothing else but the operator mixing under 
finite renormalization, see Eqs.~(\ref{physical_condensates_2a}) - (\ref{Gluon_Operator_2}). 
By means of these relations we have obtained the OPE with IR-stable Wilson coefficients which allows the limit of a vanishing light-quark mass. 

The result of an OPE in the limit $m_d \rightarrow 0$ is given by Eq.~(\ref{OPE}), which is valid for moving D-mesons 
with respect to the surrounding medium, first obtained in \cite{Hilger_Diplom}. 
The OPE in Eq.~(\ref{OPE}) is the main result of our investigation. The important case of D-mesons at rest with respect to 
the medium simplifies the OPE considerably and is given by Eqs.~(\ref{vanishing_momentum_even}) 
and (\ref{vanishing_momentum_odd}), and has first been obtained in \cite{Zschocke2}.

The OPE for a D-meson at rest has been compared with other OPE's used so far 
in the literature. Remarkably, we have found that some theoretical investigations have used seriously incorrect 
expressions for the OPE of D-mesons in matter. The aim of our investigation is, therefore, 
to present a more detailed analysis about how to obtain the correct OPE of D-mesons in medium. 
We hope that our investigation will support the correctness of prospective theoretical studies.  

\section*{Acknowledgements}

The authors would like to thank for fruitful discussions with 
Prof. Laszlo P. Csernai, Prof. Stefan Leupold, Dr. Matthias Lutz, Dr. Ronny Thomas, 
and Prof. Wolfram Weise. 
The authors are greatful to Dr. David John Broadhurst for valuable discussions and for
his great kindness in 2005 to send a copy of the Ph.D. Thesis of S.C. Generalis.
One of the authors (S.Z.) thanks for the pleasant stay at 
Helmholtz-Zentrum Dresden-Rossendorf in 2005, and for kind hospitality 
at the Department of Physics and Technology and the Bergen Computational Physics Laboratory 
of Unifob at the University of Bergen/Norway in 2006 where parts of the work were done. 
He also thanks for financial support by EU fund Hadron-Physics Integrated Infrastructure Initiative.

\begin{appendix}%================================================?

\section{Notation and conventions}\label{sct:not}

In this Appendix we briefly define the basic quantities and conventions used
throughout this work. Our expressions are obtained within the
framework of \cite{Novikov}. Contravariant four-vectors $q^{\mu} = (q_0, \ve{q})$, 
covariant four-vectors $q_{\mu} = (q_0, - \ve{q})$ in Minkowski space. The Lagrangian density of QCD reads 
\begin{eqnarray}
{\cal L}_{\rm QCD} &=& {\overline \Psi}^a\left(i\,\gamma^{\mu} \overrightarrow{D}_{\mu}^{ab} - M\,\delta^{ab} \right)\Psi^b 
-\frac{1}{4}\,G^A_{\mu\nu}\,G^{A\,\mu\nu}\,,
\label{appendix_A_1}
\end{eqnarray}

\noindent
where $M = {\rm diag} \left(m_u, m_d, m_s, m_c, m_b, m_t\right)$ is the matrix of current quark mass, 
the quark field flavors are ${\Psi} = (u,d,s,c,b,t)^{\rm T}$, and the covariant derivative 
\begin{eqnarray}
\overrightarrow{D}^{ab}_\mu(x) = \overrightarrow{\partial}_\mu {\cal I}^{ab} - i g_s {\cal A}_\mu^{ab}(x)\,,
\label{appendix_A_2}
\end{eqnarray}

\noindent
with $g_s = \sqrt{4\pi \alpha_s}$ being the strong coupling constant, and 
${\cal I}^{ab}$ is the unit matrix in color space.  
The notation $\overrightarrow{D}$ means not a vector but denotes that the partial derivative acts on the field 
to the right, vice versa $\overleftarrow{D}$ acts on the field to the left. The gluon fields are 
${\cal A}_\mu^{ab} = A_{\mu}^{A}\,t_{ab}^A$ and the gluon field strength tensor 
${\cal G}_{\mu\nu}^{ab} = G^A_{\mu\nu} t_{ab}^A$; recall the use of Einstein convention. 
Here, $t_{ab}^A=\frac{1}{2}\,\lambda^A_{ab}$ are the generators of $SU(N_c)$ where $N_c=3$ is the number 
of colors and $\lambda^A_{ab}$ are the Gell-Mann matrices. The Lorentz indices are denoted by greek indices 
$\mu,\nu = 0,1,2,3$, the color indices are denoted by small letters $a,b=1,2,3$, and the Gell-Mann index is 
denoted by a capital letter $A=1,...,8$. Finally, we note the equation of motion for quark and gluon fields:
\begin{eqnarray} 
\gamma^{\mu}\overrightarrow{D}_{\mu}\,q^a &=& - i\,m_q\,q^a\,,
\label{EOM1}
\\
\overrightarrow{D}_{AB}^{\mu}\,G_{\mu\nu}^B &=& g_s\,\sum\limits_{f} \overline{q}_f \, \gamma_{\nu}\,t^A\,q_f\,.
\label{EOM2}
\end{eqnarray}

\noindent
We also note the needed relation
\begin{eqnarray}
\overrightarrow{D}^2 \Psi &=& \left(\hat{\overrightarrow{D}} \hat{\overrightarrow{D}} 
+ \frac{1}{2}\,g_s\,\sigma\,{\cal G}\right)\Psi = \left(\frac{1}{2}\,g_s\,\sigma\,{\cal G} - m^2\right)\Psi\,,
\label{Relation_B}
\end{eqnarray}

\noindent
where $\overrightarrow{D}^2 = g^{\mu\nu}\,\overrightarrow{D}_{\mu}\,\overrightarrow{D}_{\nu}$. 
Furthermore, we use $\sigma_{\mu\nu} = \frac{i}{2}\left[\gamma_{\mu}\,,\,\gamma_{\nu}\right] = 
i\left(\gamma_{\mu}\,\gamma_{\nu} - g_{\mu\nu}\right)$, 
where the metric tensor is $g_{\mu \nu} = {\rm diag} \left(+1,-1,-1,-1\right)$, 
and for the Levi-Civita symbol we use the convention 
$\epsilon^{\alpha\mu\nu\sigma}\,\epsilon_{\alpha\mu\nu}^{\quad\;\;\tau} = 6\,g^{\sigma\tau}$ with $\epsilon_{1234} = + 1$. 

\section{Background field method in Fock-Schwinger gauge \label{Backgroundfieldmethod}}

The OPE is calculated using the background field method \cite{Reinders1} by applying Wick's theorem to the
current-current correlator and employing the Fock-Schwinger gauge 
\begin{eqnarray}
\left(x^\mu - x^{\mu}_0\right) {\cal A}_\mu (x) &=& 0 \,,
\label{gauge}
\end{eqnarray}

\noindent
first introduced in Quantum Electrodynamics \cite{Fock,Schwinger} for the photon fields, and later extended to QCD 
\cite{Fock_Schwinger_QCD,Novikov,Reinders1} for the gluon fields. Usually one choses the frame $x^{\mu}_0 = 0$.  
Very detailed aspects 
about translational invariance and the employed gauge can especially be found in \cite{Novikov}. One decisive advantage of 
Fock-Schwinger gauge is, that the gluon fields can be expressed in terms of the gluon field strength tensor:
\begin{eqnarray}
{\cal A}_{\mu}^{ab} (x) &=& \int\limits_{0}^{1}  d \alpha\,\alpha\,x^{\nu}\,{\cal G}^{ab}_{\nu\mu} (\alpha\,x)\,.
\label{Fock_Schwinger_gauge_5}
\end{eqnarray}

\noindent
This relation allows for a considerable simplification of the calculations. 
Furthermore, the Fock-Schwinger gauge allows to express partial derivatives of quark and gluon fields by covariant 
derivatives. Accordingly, a covariant expansion for gluon and quark fields in Fock-Schwinger gauge is given by 
\begin{eqnarray}
{\cal A}_\mu^{ab} (x) &=& \sum_{k=0}^{\infty}
\frac{x^\nu}{k! \left(k+2\right)} x^{\alpha_1} \ldots
x^{\alpha_k} \left( \overrightarrow{D}_{\alpha_1} \ldots \overrightarrow{D}_{\alpha_k}
{\cal G}_{\nu\mu}^{ab} \right)_{x=0} \,,
\label{eq:covariant_gluon_expansion} 
\\ 
\Psi(x) &=& \sum_{k=0}^\infty \frac{1}{k!} x^{\alpha_1} \ldots x^{\alpha_k}
\left( \overrightarrow{D}_{\alpha_1} \ldots
\overrightarrow{D}_{\alpha_k} \Psi \right)_{x=0} \,, 
\\
\overline{\Psi}(x) &=& \sum_{k=0}^{\infty} \frac{1}{k!}
x^{\alpha_1} \ldots x^{\alpha_k} \left( \bar{\Psi}
\overleftarrow{D}_{\alpha_k} \ldots \overleftarrow{D}_{\alpha_1}\right)_{x=0} \,. 
\end{eqnarray} 

\noindent
The propagator of a quark in a weak gluonic background field in coordinate space is given by 
Eq.~(\ref{eq:pert_quark_prop}).

We will now give a closed expression for the quark propagator in
momentum space which can be calculated directly from
Eq.~(\ref{eq:pert_quark_prop}) by introducing the Fourier transforms
of the quark propagator and the gluon field:
\begin{eqnarray}
S (p) &=& \int d^4 x\;{\rm e}^{i\,p\,x} \,S(x,0)\,,\quad 
\tilde{S} (p) = \int d^4 x\;{\rm e}^{ - i\,p\,x}\,S(0,x)\,,
\label{eq:fourier_propagator_1}
\\
A_\mu(p) &=& \int d^4 x\;{\rm e}^{i\,p\,x}\,A_{\mu}(x)\,. 
\label{eq:fourier_gluon}
\end{eqnarray}

\noindent
For the Fourier transform of gluon fields (\ref{eq:fourier_gluon}) we obtain 
\begin{eqnarray} 
A_\mu(p) &=& \sum_{k=0}^\infty (-i)^{k+1}
\frac{(2\pi)^4}{k!(k+2)}\left( \overrightarrow{D}_{\alpha_1}
\ldots \overrightarrow{D}_{\alpha_k} {\cal G}_{\rho\mu}(0)
\right) \left( \partial^\rho \partial^{\alpha_1} \ldots
\partial^{\alpha_k} \delta^{(4)}(p) \right),
\nonumber\\
\label{eq:gluon_field_expansion_momentum_space} 
\end{eqnarray}

\noindent
where the partial derivatives are with respect to four-momentum. 
By repeated partial integration and careful consideration of
partial derivatives acting on Dirac's delta distribution, one
can confirm the following form for the quark propagator
Eq.~(\ref{eq:pert_quark_prop}) in momentum space
\begin{eqnarray} 
S(p) &=& \sum_{k=0}^\infty S^{(k)}(p)\,, 
\label{eq:momentum_propagator} 
\\
S^{(k)}(p) &=& (-1)^k S^{(0)}(p) \underbrace{ \left( \gamma \tilde{A} \right) S^{(0)}(p) \times \,.\,.\,.\, 
\times S^{(0)}(p) \left( \gamma \tilde{A} \right) }_{\rm k}\,S^{(0)}(p)\,, 
\label{eq:propagator_sum} 
\end{eqnarray} 

\noindent
where $\tilde{A}$ is a derivative operator which naturally emerges during the Fourier transform and is defined as 
\begin{eqnarray} 
\tilde{A}_{\mu} &=& \sum_{k=0}^\infty \tilde{A}_{\mu}^{(k)}\,,
\label{eq:def_a_tilde} 
\\ 
\tilde{A}_{\mu}^{(k)} &=& 
- \frac{(-i)^{k+1} g_s}{k! (k+2)} \left(\overrightarrow{D}_{\alpha_1} \ldots
\overrightarrow{D}_{\alpha_k} {\cal G}_{\mu\nu}(0) \right) \partial^{\nu}
\partial^{\alpha_1} \ldots \partial^{\alpha_k} \,. 
\label{eq:gluon}
\end{eqnarray} 

\noindent
Due to the Fock-Schwinger gauge, the quark propagator (\ref{eq:pert_quark_prop}) does not obey translational 
invariance and, hence, is not a function of the coordinate differences and cannot be written as $S(x,y) = S(x-y)$
\cite{Novikov}. However, after performing the Fourier transformation, the difference between $S(x,0)$ and $S(0,y)$ 
is merely manifested by the operators $\tilde{A}_{\mu}$ acting on the terms to the right or to the left of them and 
the equality can be read off.

For the OPE up to mass dimension 5, only the first three terms in the sum of Eq.~(\ref{eq:momentum_propagator}) 
became relevant, and are given by ($n$ is either $c$ or $d$)
\begin{eqnarray}
S_n^{(0)} (p) &=& \frac{\hat{p} + m_n}{p^2 - m_n^2}\,\delta^{a b},
\label{Quark_Propagator_10}
\\
\nonumber\\
S_n^{(1)} (p) &=& i\,\frac{g_s}{2}\,{\cal G}_{\mu \nu}^{ab} (0)\,S_n^{(0)} (p)\,\gamma^{\mu}\,S_n^{(0)} (p)\,\gamma^{\nu}\,S_n^{(0)} (p)\,,
\label{Quark_Propagator_15}
\\
\nonumber\\
S_n^{(2)} (p) &=& \left( i \,\frac{g_s}{2}\right)^2\,{\cal G}_{\mu \nu}^{a c}(0)\,
{\cal G}_{\kappa \lambda}^{c b}(0)\,T_n^{\mu \nu \kappa \lambda} (p)\,,
\label{Quark_Propagator_20}
\end{eqnarray}

\noindent
where we have defined the tensor 
\begin{eqnarray}
T_n^{\mu \nu \kappa \lambda} (p) &=& 
S_n^{(0)} (p)\,\gamma^{\mu}\,S_n^{(0)} (p)\,\gamma^{\nu}\,S_n^{(0)} (p)\,\gamma^{\kappa}\,S_n^{(0)} (p)\,\gamma^{\lambda}
\nonumber\\
&& + S_n^{(0)} (p)\,\gamma^{\mu}\,S_n^{(0)} (p)\,\gamma^{\kappa}\,S_n^{(0)} (p)\,\gamma^{\nu}\,S_n^{(0)} (p)\,
\gamma^{\lambda}
\nonumber\\
&& + S_n^{(0)} (p)\,\gamma^{\mu}\,S_n^{(0)} (p)\,\gamma^{\kappa}\,S_n^{(0)} (p)\,\gamma^{\lambda}\,
S_n^{(0)} (p)\,\gamma^{\nu}\,.
\label{Quark_Propagator_25}
\end{eqnarray}

\section{The terms (\ref{Gluonic_Part_A}) - (\ref{Gluonic_Part_C})}\label{Gluonic_Terms}

The evaluation of the gluonic expressions (\ref{Gluonic_Part_A}) - (\ref{Gluonic_Part_C}) is involved and leads to 
intricate expressions. We therefore would like to give some basic steps about how to arrive at the final expression 
given by Eq.~(\ref{Pi_G2}), including the needed approximations. 

First, by inserting the expressions (\ref{Quark_Propagator_10}) - (\ref{Quark_Propagator_20}) into 
(\ref{Gluonic_Part_A}) - (\ref{Gluonic_Part_C}) we obtain
\begin{eqnarray}
\Pi^{G^{2,A}} (q) &=& - \left(\frac{i g_s}{2}\right)^2 \,\frac{\delta_{A B}}{2} \,\langle \Omega | : G_{\mu \nu}^A\,G_{\kappa \lambda}^B : | \Omega \rangle \; \int \frac{d^4 p}{\left(2 \pi\right)^4} \TrD   
\nonumber\\
&& \hspace{-2.0cm} \times \left( \gamma_5\,S_c^{(0)} (p) \,\gamma^{\mu}\,S_c^{(0)} (p) \,\gamma^{\nu}\, 
S_c^{(0)} (p) \gamma_5 \,S_d^{(0)} (p-q) \,\gamma^{\kappa}\,S_d^{(0)} (p-q) \,\gamma^{\lambda}\,S_d^{(0)} (p-q) \right),
\nonumber\\
\label{gluonic_terms_5}
\\
\Pi^{G^{2,B}} (q) &=& - i  \left(\frac{i g_s}{2}\right)^2 \,\frac{\delta_{A B}}{2} \,\langle \Omega | : G_{\mu \nu}^A\,G_{\kappa \lambda}^B : | \Omega \rangle 
\nonumber\\ 
&& \times \int \frac{d^4 p}{\left(2 \pi\right)^4} \TrD
\left( \gamma_5\,T_c^{\mu \nu \kappa \lambda} (p) \,\gamma_5\,S_d^{(0)} (p-q) \,\right),
\label{gluonic_terms_10}
\\
\nonumber\\
\Pi^{G^{2,C}} (q) &=& - i  \left(\frac{i g_s}{2}\right)^2 \,\frac{\delta_{A B}}{2} \,\langle \Omega | : G_{\mu \nu}^A\,G_{\kappa \lambda}^B : | \Omega \rangle 
\nonumber\\
&& \times \int \frac{d^4 p}{\left(2 \pi\right)^4} \TrD
\left( \gamma_5\,S_c^{(0)} (p+q) \,\gamma_5 \, T_d^{\mu \nu \kappa \lambda} (p)\right),
\label{gluonic_terms_15}
\end{eqnarray}

\noindent
where the trace over color-indices has been performed, i.e. $\TrD$ denotes the trace over Dirac-indices only, which are 
evaluated by means of the algebraic software HIP \cite{HIP}. Finally, we are left with the following four basic integrals:
\begin{eqnarray}
&& \int \frac{d^4 p}{(2\,\pi)^4} \frac{1}{\left[(p-q)^2 + m_d^2 \right]^n\;\left[p^2 + m_c^2\right]^k} 
\nonumber\\
&=& \frac{1}{(4\,\pi)^2}\;\frac{\Gamma(n+k-2)}{\Gamma(n)\,\Gamma(k)}\;I_{n-1,k-1,n+k-2} (q^2,m_d^2,m_c^2)\,,
\label{basic_integral_1}
\\
\nonumber\\
&& \int \frac{d^4 p}{(2\,\pi)^4} \frac{p_{\mu}}{\left[(p-q)^2 + m_d^2 \right]^n\;\left[p^2 + m_c^2\right]^k}
\nonumber\\
&=& \frac{q_{\mu}}{(4\,\pi)^2}\;\frac{\Gamma(n+k-2)}{\Gamma(n)\,\Gamma(k)}\;I_{n,k-1,n+k-2} (q^2,m_d^2,m_c^2)\,,
\label{basic_integral_2}
\\
\nonumber\\
&& \int \frac{d^4 p}{(2\,\pi)^4} \frac{p_{\mu}\,p_{\nu}}{\left[(p-q)^2 + m_d^2 \right]^n\;\left[p^2 + m_c^2\right]^k}
\nonumber\\
&=& \frac{g_{\mu\nu}}{(4\,\pi)^2}\;
\frac{1}{2}\frac{\Gamma(n+k-3)}{\Gamma(n)\,\Gamma(k)}\;I_{n-1,k-1,n+k-3} (q^2,m_d^2,m_c^2)
\nonumber\\
&& + \frac{q_{\mu}\,q_{\nu}}{(4\,\pi)^2}\;\frac{\Gamma(n+k-2)}{\Gamma(n)\,\Gamma(k)}\;I_{n+1,k-1,n+k-2} (q^2,m_d^2,m_c^2)\,,
\label{basic_integral_3}
\\
\nonumber\\
&& \int \frac{d^4 p}{(2\,\pi)^4} 
\frac{p_{\mu}\,p_{\nu}\,p_{\kappa}}{\left[(p-q)^2 + m_d^2 \right]^n\;\left[p^2 + m_c^2\right]^k}
\nonumber\\
&=& \frac{g_{\mu\nu}\,q_{\kappa} + g_{\mu\kappa}\,q_{\nu} + g_{\nu\kappa}\,q_{\mu}}{(4\,\pi)^2}\;\frac{1}{2}
\frac{\Gamma(n+k-3)}{\Gamma(n)\,\Gamma(k)}\;I_{n,k-1,n+k-3} (q^2,m_d^2,m_c^2)
\nonumber\\
&& + \frac{q_{\mu}\,q_{\nu}\,q_{\kappa}}{(4\,\pi)^2}\;
\frac{\Gamma(n+k-2)}{\Gamma(n)\,\Gamma(k)}\;I_{n+2,k-1,n+k-2} (q^2,m_d^2,m_c^2)\,,
\label{basic_integral_4}
\end{eqnarray}

\noindent
where $\Gamma(x) = \int\limits_{0}^{\infty} dt \,t^{x-1}\,{\rm e}^{-t}$ is the gamma function, 
and the master integral reads 
\begin{eqnarray}
I_{i,j,k} (q^2,m_d^2,m_c^2) &=& \int \limits_0^1 d \alpha\,
\frac{\alpha^i\,\left(1 - \alpha\right)^j}{\left[\alpha \left(1 - \alpha\right)\,q^2 + \alpha\,m_d^2 
+ \left(1 - \alpha\right)\,m_c^2\right]^k}\;.
\label{basic_integral_5}
\end{eqnarray}

\noindent
Needless to say that these integrals lead to rather cumbersome expressions for finite mass $m_d$ and $m_c$. 
On the other side it is not possible to put them to zero since the integrals are IR-divergent, that means 
they contain terms like $m_d^{-1}$ and $\ln m_d$. In order to extract the corresponding terms $\propto m_d^{-1}$ one has to 
multiply the integral with the lowest power $n$ of $m_d$ that gives a finite result for the mass going to zero. What 
remains is the corresponding coefficient for the $m_d^{-n}$ term. All the other coefficients can be obtained by taking 
suitable derivatives of $m_d^n I_{ijk}$ with respect to $m_d$ and setting afterward $m_d$ equal to zero. The terms 
$\propto \ln m_d$ can be obtained by taking the first derivative of $I_{ijk}$ and extracting the term 
$\propto m_d^{-1}$ from that expression in the same way. In this respect we also note the useful derivative relations 
\begin{eqnarray}         
\frac{\partial}{\partial m_d^2} I_{ijk}(q^2, m_d^2, m_c^2) &=& -k I_{i+1,j,k+1}(q^2, m_d^2, m_c^2)\,,
\label{derivative_1}
\\         
\frac{\partial}{\partial m_c^2} I_{ijk}(q^2, m_d^2, m_c^2) &=& -k I_{i,j+1,k+1}(q^2, m_d^2, m_c^2)\,,
\label{derivative_2}
\\
\frac{\partial}{\partial q^2} I_{ijk}(q^2, m_d^2, m_c^2) &=& -k I_{i+1,j+1,k+1}(q^2, m_d^2, m_c^2)\,.
\label{derivative_3}
\end{eqnarray} 

\noindent
Sometimes, some additional manipulations have to be made in order to obtain meaningful expressions. Using
\begin{eqnarray}
\arctan z &=& \frac{1}{2i} \ln \frac{1+iz}{1-iz}
\end{eqnarray}

\noindent
one gets
\begin{eqnarray}
\arctan \left( i \frac{q^2-m_d^2}{q^2+m_d^2} \right) &=& \frac{1}{2i} \ln \frac{m_d^2}{q^2} \;,
\end{eqnarray}

\noindent
which is a source of mass logarithms. Another source of terms $\propto \ln m_d$ arises from a slightly different 
expression, namely
\begin{eqnarray}
\arctan \left( i \frac{q^2+m_c^2-m_d^2}{\sqrt{2q^2 m_d^2 + q^4 + 2q^2 m_c^2 + m_c^4 - 2 m_c^2 m_d^2 + m_d^4}} \right)
\stackrel{m_d=0}{\rightarrow} \arctan (i)\,,
\end{eqnarray}

\noindent
which is not well defined. Expanding the fraction in $m_d$ and keeping only the lowest power, which is the dominant 
contribution for $m_d \rightarrow 0$, leads to
\begin{eqnarray}
\arctan \left( i \frac{q^2+m_c^2-m_d^2}{\sqrt{2q^2 m_d^2 + q^4 + 2q^2 m_c^2 + m_c^4 - 2 m_c^2 m_d^2 + m_d^4}} \right) 
\stackrel{m_d \approx 0}{=} \frac{1}{2i} \ln \frac{q^2 m_d^2}{(q^2+m_c^2)^2}\,.
\nonumber\\
\end{eqnarray}

\noindent
This term arises in the tensor part of the term (\ref{gluonic_terms_15}) and causes an infrared divergent Wilson 
coefficient. It prohibits us from taking the limit $m_d \rightarrow 0$ at this stage. These formulas and the technique 
described here in detail allow us to evaluate the integrals (\ref{gluonic_terms_5}) - (\ref{gluonic_terms_15}). We 
underline that the expressions (\ref{gluonic_terms_5}) - (\ref{gluonic_terms_15}) are symmetric in exchanging charm and 
down quarks. However, it is obvious that the needed series expansions in respect to the small quark mass $m_d$ destroy this 
symmetry. We finally have obtained the following results:
\begin{eqnarray}
\Pi^{G^2,A}_{\rm scalar} (q) &=& - \frac{1}{8} \GGV \frac{1}{q^2 - m_c^2} \,,
\label{scalar_A}
\\
\nonumber\\
 \Pi^{G^2,B}_{\rm scalar} (q) &=& + \frac{1}{24} \GGV \frac{1}{q^2 - m_c^2} \,,
\label{scalar_B}
\\
\nonumber\\
\Pi^{G^2,C}_{\rm scalar} (q) &=& \GGV \left( - \frac{1}{12} \frac{m_c}{m_d} \frac{1}{q^2 - m_c^2} 
+ \frac{1}{24} \frac{1}{q^2 - m_c^2} - \frac{1}{24} \frac{m_c^2}{( q^2 - m_c^2)^2} \right),
\nonumber\\
\label{scalar_C}
\end{eqnarray}

\noindent
where $G^2 = G_{\mu\nu}^A\,G_{\mu\nu}^A$. 
We recognize that the term $\propto m_d^{-1}$ prohibits us from regarding one quark to be massless in evaluating the 
scalar terms. For the tensor part we obtain for $m_d \approx 0$ the following results:  
\begin{eqnarray}
\Pi^{G^2,A}_{\rm tensor} (q) &=& \GGM \left( q^2 - 4 \frac{(vq)^2}{v^2} \right)
\nonumber\\
&& \times \left( - \frac{1}{3} \frac{m_c^2}{q^6} \ln \left( - \frac{m_c^2}{q^2 - m_c^2} \right)
- \frac{1}{6q^2} \frac{1}{q^2 - m_c^2} + \frac{1}{3q^4} \right)\, ,
\label{tensor_A}
\\
\nonumber\\
\Pi^{G^2,B}_{\rm tensor} (q) &=& \GGM \left( q^2 - 4\frac{(vq)^2}{v^2} \right)
\left( - \frac{1}{9q^4} - \frac{1}{18q^2} \frac{1}{q^2 - m_c^2} \right.
\nonumber\\
&& \left. - \frac{1}{9q^4} \ln\left( - \frac{m_c^2}{q^2-m_c^2}\right)
+ \frac{m_c^2}{9q^6} \ln\left( - \frac{m_c^2}{q^2-m_c^2}\right)\right)\,,
\label{tensor_B}
\\
\nonumber\\
\Pi^{G^2,C}_{\rm tensor} (q) &=& \GGM \left( q^2 - 4\frac{(vq)^2}{v^2} \right)
\left( - \frac{2}{9q^4} + \frac{1}{18q^2} \frac{1}{q^2 - m_c^2} \right.
\nonumber\\
&& - \frac{1}{6q^2} \frac{m_c^2}{(q^2 - m_c^2)^2}
+ \left( \frac{1}{9q^4}
+ \frac{2m_c^2}{9q^6} \right)
\ln \left(- \frac{m_c^2}{q^2 - m_c^2} \right)
\nonumber\\
&& - \frac{1}{9q^2} \left( \frac{m_c^2}{(q^2 - m_c^2)^2} + \frac{1}{q^2 - m_c^2} \right)
\ln \left( \frac{m_d^2}{m_c^2} \right)
\nonumber\\
&& \left. - \frac{2}{9q^2} \left( \frac{m_c^2}{(q^2 - m_c^2)^2} + \frac{1}{q^2 - m_c^2} \right)
\ln \left(- \frac{m_c^2}{q^2 - m_c^2} \right)\right).
\label{tensor_C}
\end{eqnarray}

\noindent
Here, we observe the occurrence of a term $\propto \ln m_d$, which again prohibits us from taking the limit 
$m_d \rightarrow 0$. The divergent terms cancel, however, after the introduction of physical condensates. 
According to Eq.~(\ref{Gluonic_Part}), we collect all the 
terms (\ref{scalar_A}) - (\ref{tensor_C}) and obtain the expression given by Eq.~(\ref{Pi_G2}). 

\section{Proof of relation (\ref{physical_condensates_2d}) \label{Appendix_Mixing}}

The operator in relation (\ref{physical_condensates_2d}) reads ${\cal O} \left[ \overrightarrow{D}_{\mu} \right] = 
i\,\overrightarrow{D}_{\mu}\,i\,\overrightarrow{D}_{\nu}$. 
From the operator mixing (\ref{definition_physical_condensates}) we obtain in one-loop approximation 
\begin{eqnarray}
\langle \Omega | \bar{q}\,i\,\overrightarrow{D}_{\mu}\,i\,\overrightarrow{D}_{\nu}\,q | \Omega \rangle^{(1)} &=&
\langle \Omega | : \bar{q} \,i\,\overrightarrow{D}_{\mu}\,i\,\overrightarrow{D}_{\nu}\,q : | \Omega \rangle^{(0)} 
\nonumber\\
&& \hspace{-2.0cm} - i \int \frac{d^4p}{(2\pi)^4}
\langle \Omega | : \TrCD \left[ \left(p_\mu + \tilde{A}_\mu\right) \left(p_\nu + \tilde{A}_\nu\right) S(p)\right] : 
| \Omega \rangle^{(0)}\,.
\label{appendix_mixing_10}
\end{eqnarray}

\noindent
In consistence with the whole approach, we determine the integral up to order ${\cal O} (g^2)$,  
and the expression in the last line can be separated into three terms as follows: 
\begin{eqnarray}
T_1 &=& - i \int\frac{d^4p}{(2\pi)^4}\,\langle\Omega | : \TrCD \left[\tilde{A}^{(0)}_\mu \,\tilde{A}^{(0)}_\nu\,
S^{(0)}(p)\right] : |\Omega\rangle^{(0)}\,,
\label{appendix_mixing_A}
\\
T_2 &=& - i\int\frac{d^4p}{(2\pi)^4}\,p_{\mu}\,p_{\nu}\,\langle\Omega| : \TrCD \left[S^{(0)}(p)\right] : |\Omega\rangle^{(0)}\,,
\label{appendix_mixing_B}
\\
T_3 &=& - i \int \frac{d^4p}{(2\pi)^4} p_{\mu}\,p_{\nu}\,
\langle \Omega | : \TrCD \left[ S^{(2)} (p)\right] : | \Omega \rangle^{(0)}\,,
\label{appendix_mixing_C}
\end{eqnarray}

\noindent
while all other terms vanish. 
Let us consider explicitly $T_1$. Using the expressions of the fermion propagator 
given by Eqs.~(\ref{Quark_Propagator_10}) - (\ref{Quark_Propagator_20}), and the derivative operator 
$\displaystyle \tilde{A}^{(0)}_\mu = \frac{i\,g_s}{2}\,{\cal G}_{\mu\nu}\, \partial^{\nu}$ which acts on the 
quark propagators only, we obtain 
\begin{eqnarray}
T_1 &=& - i \left(\frac{i\,g_s}{2}\right)^2 
\langle\Omega | : G_{\mu \kappa}^A \, G_{\nu \lambda}^B : | \Omega\rangle^{(0)}\;
\TrC\left[\frac{\lambda^A}{2}\;\frac{\lambda^B}{2}\right]
\nonumber\\
&& \times \mu^{2 \epsilon} \int\frac{d^D p}{(2\pi)^D}\,\frac{1}{\left(p^2 - m_q^2\right)^3} 
\TrD \left[\left(\hat{p} + m_q\right) \gamma_{\kappa}\left(\hat{p} + m_q\right) \gamma_{\lambda} 
\left(\hat{p} + m_q\right)\right] + \kappa \leftrightarrow \lambda \;,
\nonumber\\
\label{appendix_mixing_15}
\end{eqnarray}

\noindent
where the integrals are divergent and have to be evaluated in $D = 4 - 2\,\epsilon$ dimensions, 
i.e. by means of dimensional regularization; $\mu$ is the renormalization scale of finite renormalization 
governed by the renormalization group equation \cite{Zuber,Muta}. 

Now we insert the projection (\ref{projection2b}) and perform the Dirac traces; 
note the color trace is $\TrC \lambda^A\,\lambda^B = 2\,\delta^{AB}$. 
After a lengthly but straigtforward calculation we finally obtain 
\begin{eqnarray}
T_1 &=& - i \left(\frac{i\,g_s}{2}\right)^2 \frac{4}{96} \langle \Omega | : G^2 : | \Omega \rangle^{(0)} 
\nonumber\\
&& \times \mu^{2 \epsilon} \int\frac{d^D p}{(2\pi)^D}\,\frac{1}{\left(p^2 - m_q^2\right)^3} 
\left(24\,m_q^3\,g_{\mu\nu} + 8\,m_q\,p^2\,g_{\mu\nu} - 32\,m_q\,p_{\mu}p_{\nu}\right)
\nonumber\\
\nonumber\\
&& + i \left(\frac{i\,g_s}{2}\right)^2 \frac{4}{24} 
\langle \Omega | : \left( \frac{\left(v\,G\right)^2}{v^2} - \frac{G^2}{4} \right) : | \Omega \rangle^{(0)}
\nonumber\\
&& \times \mu^{2 \epsilon} \int\frac{d^D p}{(2\pi)^D}\,\frac{1}{\left(p^2 - m_q^2\right)^3}
\left(24\,m_q^3\,g_{\mu\nu} + 8\,m_q\,p^2\,g_{\mu\nu} - 32\,m_q\,p_{\mu}p_{\nu}\right)
\nonumber\\
\nonumber\\
&& - i \left(\frac{i\,g_s}{2}\right)^2 \frac{8}{24} 
\langle \Omega | : \left( \frac{\left(v\,G\right)^2}{v^2} - \frac{G^2}{4} \right) : | \Omega \rangle^{(0)}
\nonumber\\
&& \times \mu^{2 \epsilon} \int\frac{d^D p}{(2\pi)^D}\,\frac{1}{\left(p^2 - m_q^2\right)^3}
\bigg( - 8\,m_q (p^2 - m_q^2) g_{\mu\nu} + 16\,m_q\,p^2\,v_{\mu} v_{\nu} + 16\,m_q^3\,v_{\mu}v_{\nu}
\nonumber\\
&& \quad \quad -32\,m_q\,(p\,v) (v_{\mu} p_{\nu} + v_{\nu} p_{\mu}) + 32\,m_q (p\,v)^2 g_{\mu\nu}\bigg), 
\label{appendix_mixing_20}
\end{eqnarray}

\noindent
where we have kept some ratios in order to show more explicitly their source from the projection. The two needed integrals 
can be evaluated by standard techniques and in $\overline{\rm MS}$ scheme \cite{MS_bar} they are given by: 
\begin{eqnarray}
I_1 &=& \int\frac{d^4 p}{(2\pi)^4}\,\frac{1}{\left(p^2 - m_q^2\right)^3} 
= - \frac{1}{2}\,\frac{1}{m_q^2}\,\frac{i}{(4\,\pi)^2}\,,
\label{intergal_1}
\\
I_2 &=& \mu^{2 \epsilon}\int\frac{d^D p}{(2\pi)^D}\,\frac{p_{\mu} p_{\nu}}{\left(p^2 - m_q^2\right)^3} = 
- \frac{1}{4}\,g_{\mu\nu} \frac{i}{(4\,\pi)^2}\,\left[\frac{1}{2} - \ln \frac{\mu^2}{m_q^2}\right]\,.
\label{intergal_2}
\end{eqnarray}

\noindent
The second integral is logarithmically divergent and has to be evaluated by means of dimensional regularization. 
Inserting both integrals into (\ref{appendix_mixing_20}), we arrive at 
\begin{eqnarray}
T_1 &=& + \frac{1}{32}\,m_q\,g_{\mu\nu}\,\langle \Omega | : \frac{\alpha_s}{\pi}\,G^2 : | \Omega \rangle^{(0)} 
\nonumber\\
&& - \frac{1}{24}\,m_q\,g_{\mu\nu}\,
\langle\Omega|:\frac{\alpha_s}{\pi} \left(\frac{\left(v\,G\right)^2}{v^2} - \frac{G^2}{4}\right):|\Omega\rangle^{(0)}
\nonumber\\
&& + \frac{1}{6} \,m_q\,v_{\mu}\,v_{\nu}\,
\langle\Omega|:\frac{\alpha_s}{\pi} \left(\frac{\left(v\,G\right)^2}{v^2} - \frac{G^2}{4}\right):|\Omega\rangle^{(0)}\,.
\label{appendix_mixing_25}
\end{eqnarray}

\noindent
Now we consider the term $T_2$. Using $\langle \Omega | \Omega \rangle = 1$, and inserting (\ref{Quark_Propagator_10}) 
into (\ref{appendix_mixing_B}) yields 
\begin{eqnarray}
T_2 &=& - \,3\,\,m_q\;g_{\mu\nu}\,i \, \mu^{2\,\epsilon}\int\frac{d^D p}{(2\pi)^D}\,\frac{p^2}{p^2-m_q^2}\,,
\label{appendix_mixing_26}
\end{eqnarray}

\noindent
where we have used that the only possible Lorentz structure is the metric tensor, and we have used 
$\TrC \left[\delta^{a b} \right] = 3$, $\TrD \left[\hat{p}\right] = 0$, $\TrD \left[m_q\right] = D\,m_q$ and 
$g_{\mu}^{\mu} = D$, where $D=4 - 2\,\epsilon$.  
The nominator of the integral is treated by means of $p^2 = (p^2 - m_q^2) + m_q^2$, and then using the common law 
$\int\frac{d^D p}{(2\pi)^D} = 0$ by definition, e.g. \cite{Integral}, we conclude the superficial degree of divergence 
of this integral is $2$. Standard evaluation in $\overline{\rm MS}$ scheme \cite{MS_bar} yields:
\begin{eqnarray}
I_3 &=& \mu^{2\,\epsilon}\int\frac{d^D p}{(2\pi)^D}\,\frac{m_q^2}{p^2-m_q^2} = \frac{i}{(4\,\pi)^2}\;m_q^4\; 
\left(\ln \frac{\mu^2}{m_q^2} + 1\right).
\label{appendix_mixing_27}
\end{eqnarray}

\noindent
Thus, we obtain 
\begin{eqnarray}
T_2 &=& \frac{3\,m_q^5}{16\,\pi^2}g_{\mu\nu} \left( \ln{\frac{\mu^2}{m_q^2}} + 1 \right).
\label{appendix_mixing_28}
\end{eqnarray}

\noindent
The determination of $T_3$ is a lenghtly calculation because of the more involved Dirac traces, so we 
present just the final result. An application of the same techniques as used by $T_1$ yields for $T_3$: 
\begin{eqnarray}
T_3 &=& - \left(\frac{5}{96} - \frac{1}{16}\,\ln \frac{\mu^2}{m_q^2} \right) m_q\,g_{\mu\nu}
\langle \Omega | : \frac{\alpha_s}{\pi}\,G^2 : | \Omega \rangle^{(0)}
\nonumber\\
&& + \left(\frac{5}{216} - \frac{1}{36}\,\ln \frac{\mu^2}{m_q^2} \right) m_q\,g_{\mu\nu}
\langle\Omega|:\frac{\alpha_s}{\pi}\left(\frac{\left(v\,G\right)^2}{v^2} - \frac{G^2}{4}\right):|\Omega\rangle^{(0)}
\nonumber\\
&& - \left(\frac{5}{54} - \frac{1}{9}\,\ln \frac{\mu^2}{m_q^2} \right) m_q\,v_{\mu}\,v_{\nu}\,
\langle\Omega|:\frac{\alpha_s}{\pi} \left(\frac{\left(v\,G\right)^2}{v^2} - \frac{G^2}{4}\right):|\Omega\rangle^{(0)}\,.
\label{appendix_mixing_40}
\end{eqnarray}

\noindent
According to Eq.~(\ref{appendix_mixing_10}) we add the terms $T_1$, $T_2$ and $T_3$ and finally obtain: 
\begin{eqnarray}
\langle | \Omega \bar{q} i \overrightarrow{D}_{\mu} i \overrightarrow{D}_{\nu} q | \Omega \rangle^{(1)} &=& 
\langle \Omega | : \bar{q} i \overrightarrow{D}_{\mu} i \overrightarrow{D}_{\nu} q : | \Omega \rangle^{(0)}
+ \frac{3\,m_q^5}{16\,\pi^2}g_{\mu\nu} \left( \ln{\frac{\mu^2}{m_q^2}} + 1 \right)
\nonumber\\
&& \hspace{-3.5cm}
+ \frac{m_q}{16}g_{\mu\nu} \left( \ln{\frac{\mu^2}{m_q^2}} - \frac{1}{3} \right) \langle \Omega | : \frac{\alpha_s}{\pi}\;G^2 : |\Omega\rangle^{(0)}
\nonumber\\
&& \hspace{-3.5cm}
- \frac{m_q}{36} \left( g_{\mu\nu} - 4\frac{v_\mu v_\nu}{v^2} \right) \left( \ln{\frac{\mu^2}{m_q^2}} + \frac{2}{3} \right)
\langle \Omega | : \frac{\alpha_s}{\pi}\left(\frac{\left(v G\right)^2}{v^2} - \frac{G^2}{4}\right) : |\Omega\rangle^{(0)}\,,
\label{appendix_mixing_45}
\end{eqnarray}

\noindent
which is just relation (\ref{physical_condensates_2d}).

\section{Projections}\label{projections}

Condensates are vacuum or in-medium expectation values of quantum field operators. They reflect basic properties of the 
ground state of QCD or of the medium. Condensates are assumed to be color singlets, Lorentz invariants and invariants 
under parity transformations and time reversal, according to the corresponding symmetries assumed for the medium. 
Therefore, one has to project out color, spinor and Lorentz indices from several structures that appear 
during our calculations. Expectation values which are not invariant under parity transformations and time reversal 
are supposed to be zero. In what follows we adopt the method described in \cite{Ji93}. 

\subsection{Projection of color and Dirac structure}

Up to mass dimension 5 we meet the following structures
\begin{eqnarray}
&&\langle \Omega | \bar{q}^{\,a}_i q^b_j | \Omega \rangle \, , \quad
\langle \Omega | \left( \bar{q}_i \overrightarrow{D}_\mu \right)^a q^b_j | \Omega \rangle \,, \quad
\langle \Omega | \left( \bar{q}_i \overrightarrow{D}_\mu \overrightarrow{D}_\nu \right)^a q^b_j |\Omega \rangle \,,
\\
&& \langle \Omega | G_{\mu\nu}^A G_{\kappa\lambda}^B | \Omega \rangle\,, \quad 
\langle \Omega | \bar{q}_i^{\,a} {\cal G}_{\mu\nu}^{ab} q^b_j | \Omega \rangle\,,
\end{eqnarray}

\noindent
where the Dirac indices $i,j=0,1,2,3$. 
The last structure is already invariant under color rotations, thus one does not need to take care of color projections for 
this one. One can expand the remaining other condensates using an orthogonal basis. For $N_c \times N_c$ dimensional 
matrices such a set is given by the generators $t^A_{ab}$ of $SU(N_c)$ supplemented by the unit matrix ${\cal I}$ in 
$N_c$ dimensions; an appropriate scalar product is given by the trace operation $(A,B) = {\rm Tr} (AB)$. Using $SU(3)$  
generator identities the prescription for the expansion in terms of the generators of SU($N_c$) reads as follows
\begin{eqnarray}
M_{ab} &=& \sum_A c_A t^A_{ab} + c_{\cal I} {\cal I}_{ab} \;,
\\
c_A &=& \frac{2}{\TrC (t^A t^A)} \TrC (t^A M) \quad {\rm and} \quad
c_{\cal I} = \frac{1}{N_c} \TrC (M)\,.
\end{eqnarray}

\noindent
In order to retain color singlets the only non-vanishing expansion coefficient is $c_{\cal I}$. All the other 
coefficients belong to expectation values which do not transform as scalars under color transformations.

The projection of spinor indices proceeds in an analog way. A complete set is given by the elements $O_k$ of the Clifford 
algebra, i.e.\ $O_k \in \left\{{\cal I}, \gamma_\mu, \sigma_{\mu\nu}, i \gamma_5 \gamma_\mu, \gamma_5 \right\}$, 
satisfying $\TrD \left(O_i\,O^j\right) = 4 \delta_i^j$. The expansion reads
\begin{eqnarray}
\langle \Omega | \bar{q}_i {\cal O}_{\mu\nu \ldots} q_j | \Omega \rangle &=& \sum_k d_k O^k_{ji}\,, 
\quad d_k = \frac{1}{4} \langle \Omega | \bar{q} O_k {\cal O}_{\mu\nu \ldots} q | \Omega \rangle\,,
\end{eqnarray}

\noindent
resulting in
\begin{eqnarray}
\langle \Omega | \bar{q}_i {\cal O}_{\mu\nu \ldots} q_j | \Omega \rangle &=& 
\frac{1}{4} \bigg( \langle \Omega | \bar{q} {\cal O}_{\mu\nu \ldots} q | \Omega \rangle {\cal I}_{ji} 
+ \langle \Omega | \bar{q} \gamma_\mu {\cal O}_{\mu\nu \ldots} q | \Omega \rangle \gamma^\mu_{ji} 
\nonumber\\
&& + \frac{1}{2} \langle \Omega | \bar{q} \sigma_{\mu\nu} {\cal O}_{\mu\nu \ldots} q | 
\Omega \rangle \sigma^{\mu\nu}_{ji} 
 - \langle \Omega | \bar{q} \gamma_5 \gamma_\mu {\cal O}_{\mu\nu \ldots} q | \Omega \rangle \gamma_5\gamma^\mu_{ji}
\nonumber\\
&& + \langle \Omega | \bar{q} \gamma_5 {\cal O}_{\mu\nu \ldots} q | \Omega \rangle \gamma_{5,ji} \bigg).
\end{eqnarray}

\noindent
The notation used is: $\gamma_5\gamma^\mu_{ji} \equiv (\gamma_5\gamma^\mu)_{ij}$ and 
$\gamma_{5,ji} \equiv (\gamma_5)_{ij}$. Together with color projection one gets
\begin{eqnarray}
\langle \Omega | \bar{q}^a_i {\cal O}_{\mu\nu \ldots} q^b_j | \Omega \rangle 
&=& \frac{\delta^{ab}}{12} \bigg( \langle \Omega | \bar{q} {\cal O}_{\mu\nu \ldots} q | \Omega \rangle {\cal I}_{ji}
+ \langle \Omega | \bar{q} \gamma_\mu {\cal O}_{\mu\nu \ldots} q | \Omega \rangle \gamma^\mu_{ji}
\nonumber\\
&& + \frac{1}{2} \langle \Omega | \bar{q} \sigma_{\mu\nu} {\cal O}_{\mu\nu \ldots} q | 
\Omega \rangle \sigma^{\mu\nu}_{ji} 
- \langle \Omega | \bar{q} \gamma_5 \gamma_\mu {\cal O}_{\mu\nu \ldots} q | \Omega \rangle \gamma_5\gamma^\mu_{ji} 
\nonumber\\
&& + \langle \Omega | \bar{q} \gamma_5 {\cal O}_{\mu\nu \ldots} q | \Omega \rangle \gamma_{5,ji} \bigg).
\end{eqnarray}

\noindent
Here ${\cal O}_{\mu\nu \ldots}$ stands for an arbitrary operator with Lorentz indices $\mu\nu \cdots$. In Ref.~\cite{Ji93} 
it has been stated that terms corresponding to the projection onto $\sigma_{\mu\nu}, \gamma_5 \gamma_\mu, \gamma_5$ do 
not appear due to parity and/or time reversal. This statement can be misunderstood. It is true for 
$\gamma_5 \gamma_\mu, \gamma_5$, however not for $\sigma_{\mu\nu}$, which does not contribute because there is no 
independent Lorentz structure that reflects the symmetry properties of $\sigma_{\mu\nu}$. In fact, it contributes if the 
gluon field strength tensor enters the operator product (i.e. ${\cal O}_{\mu\nu} = {\cal G}_{\mu\nu}$).

\subsection{Projection of Lorentz structure}

A striking difference of vacuum and in-medium projections appears when projecting Lorentz indices. It is important to note 
that this projection is not an expansion in terms of a complete orthogonal set. It strongly depends on the structures 
available to perform the projection. In vacuum, there are only two independent objects, the metric tensor $g_{\mu\nu}$ and 
the total antisymmetric symbol $\epsilon_{\mu\nu\alpha\beta}$, being a pseudo-tensor under parity transformations. In 
medium the condensates also depend on the medium's four-velocity $v^\mu$ which, therefore, is an additional structure for 
projections. As a result, also pseudo-vectorial structures can be invariant under parity transformations.

Now we give a list of the in-medium projections up to mass dimension 5. Terms that violate time reversal or parity 
invariance are omitted. Equations of motion enable us to rewrite the condensates in terms of canonical 
condensates \cite{Gr94}.

\subsubsection{Condensates of mass dimension 3}

The in-medium projections of the Lorentz structure of the condensates of mass dimension 3 read:
\begin{eqnarray}
\langle \Omega | \bar{q} q | \Omega \rangle &=& \langle \Omega | \bar{q} q | \Omega \rangle \,,
\\
\langle \Omega | \bar{q} \gamma_\mu q | \Omega \rangle &=& 
\langle \Omega | \bar{q} \hat{v} q | \Omega \rangle \frac{v_\mu}{v^2} \,.
\end{eqnarray}

\noindent
A condensate of the type $\langle \bar{q} \sigma_{\mu\nu} q \rangle$ occurs neither in vacuum nor in a medium since 
there is no possibility to create an antisymmetric structure in the Lorentz indices $\mu,\nu$. Condensates of the type 
$\langle \bar{q} \gamma_5 \gamma_\mu q \rangle, \langle \bar{q} \gamma_5 q \rangle$ can not be projected onto 
structures that are invariant under parity transformations.

\subsubsection{Condensates of mass dimension 4}

The in-medium projections of the Lorentz structure of the condensates of mass dimension 4 read:
\begin{eqnarray}
\langle \Omega | \bar{q} \overrightarrow{D}_\mu q | \Omega \rangle &=& 
-\langle \Omega | \bar{q} \hat{v} q | \Omega \rangle \,i \frac{m_q v_\mu}{v^2} \,,
\\
\langle \Omega | \bar{q} \gamma_\mu \overrightarrow{D}_\nu q | \Omega \rangle &=& 
-\langle \Omega | \bar{q} q | \Omega \rangle \, i \frac{m_q}{4} g_{\mu\nu} 
\nonumber\\
&& \hspace{-2.5cm} - \left[ i \frac{m_q}{4} \langle \Omega | \bar{q} q | \Omega \rangle
+ \langle \Omega | \bar{q} \hat{v} \frac{(vD)}{v^2} q | \Omega \rangle \right]
\frac{1}{3} \left( g_{\mu\nu} - 4 \frac{v_\mu v_\nu}{v^2} \right),
\label{projection2a}
\\
\langle \Omega | G_{\mu\nu}^A G_{\kappa\lambda}^B | \Omega \rangle &=& \frac{\delta^{AB}}{96}
\left( g_{\mu\kappa} g_{\nu\lambda} - g_{\mu\lambda} g_{\nu\kappa} \right) \langle \Omega | G^2 | \Omega \rangle 
\nonumber\\
&& - \frac{\delta^{AB}}{24} 
\langle \Omega | \left(\frac{(v G)^2}{v^2} - \frac{G^2}{4}\right) | \Omega \rangle
 {\rm S}_{\mu\nu\kappa\lambda} \,,
\label{projection2b} 
\end{eqnarray}

\noindent
where we have defined
\begin{eqnarray}
{\rm S}_{\alpha\beta\mu\nu} &=& \left( g_{\alpha\mu} g_{\beta\nu} - g_{\alpha\nu} g_{\beta\mu}
- 2 \left( g_{\alpha\mu} \frac{v_\beta v_\nu}{v^2} - g_{\alpha\nu} \frac{v_\beta v_\mu}{v^2}
+ g_{\beta\nu} \frac{v_\alpha v_\mu}{v^2} - g_{\beta\mu} \frac{v_\alpha v_\nu}{v^2} \right) \right).
\nonumber\\
\label{S_tensor}
\end{eqnarray}

\noindent
Again, terms of the form $\langle \bar{q}\gamma_5\gamma_\mu \overrightarrow{D}_\mu q \rangle, \langle \bar{q}\gamma_5 
\overrightarrow{D}_\mu q \rangle$ 
do not have a projection due to the requirement of parity invariance. The same holds true for 
$\langle \bar{q} \sigma_{\mu\nu} \overrightarrow{D}_\kappa q \rangle$, which can be contracted with 
$\epsilon_{\mu\nu\kappa\lambda} v^\lambda$ giving an odd term with respect to parity.

\subsubsection{Condensates of mass dimension 5}

The in-medium projections of the Lorentz structure of the condensates of mass dimension 5 read:
\begin{eqnarray}
&&\langle \Omega | \bar{q} \overrightarrow{D}_\mu \overrightarrow{D}_\nu q | \Omega \rangle = 
-\langle \Omega | \bar{q} q | \Omega \rangle \frac{m_q^2}{4} g_{\mu\nu}
+ \langle \Omega | \bar{q} g_s \sigma {\cal G} q | \Omega \rangle \frac{1}{8} g_{\mu\nu}
\nonumber\\
&& \quad - \left[ \frac{m_q^2}{4} \langle \Omega | \bar{q} q | \Omega \rangle
- \frac{1}{8} \langle \Omega | \bar{q} g_s \sigma {\cal G} q | \Omega \rangle
+ \langle \Omega | \bar{q} \frac{(vD)^2}{v^2} q | \Omega \rangle \right]
\nonumber\\
&& \quad \quad \times \frac{1}{3} \left( g_{\mu\nu} - 4 \frac{v_\mu v_\nu}{v^2} \right),
\label{projection3a}
\end{eqnarray}

\begin{eqnarray}
&& \langle \Omega | \bar{q} \gamma_\mu \overrightarrow{D}_\nu \overrightarrow{D}_\alpha q | \Omega  \rangle
\nonumber\\
&& \quad = \frac{1}{v^4} \langle \Omega | \bar{q} \hat{v} (vD)^2 q | \Omega \rangle
\left( \frac{2\,v_\mu v_\nu v_\alpha}{v^2}
- \frac{1}{3} \left( v_\mu g_{\nu\alpha} + v_\nu g_{\mu\alpha} + v_\alpha g_{\mu\nu} \right) \right)
\nonumber\\
&& \quad - \frac{1}{6\,v^2} \langle \Omega | \bar{q} \hat{v} g_s \sigma {\cal G} q | \Omega \rangle
\left( \frac{v_\mu v_\nu v_\alpha}{v^2} - v_\mu g_{\nu\alpha} \right)
\nonumber\\
&& \quad + \frac{m_q^2}{3\,v^2} \langle \Omega | \bar{q} \hat{v} q | \Omega \rangle
\left( \frac{v_\mu v_\nu v_\alpha}{v^2} - v_\mu g_{\nu\alpha} \right)
\nonumber\\
&& \quad + \frac{i m_q}{3\,v^2} \langle \Omega | \bar{q} (vD) q | \Omega \rangle
 \left( \frac{2 v_\mu v_\nu v_\alpha}{v^2} - v_\nu g_{\mu\alpha} - v_\alpha g_{\mu\nu} \right),
\label{dim_5_A}
\end{eqnarray}

\begin{eqnarray}
&& \langle \Omega | \bar{q} \gamma_5 \gamma_\alpha {\cal G}_{\mu\nu} q | \Omega \rangle = - \frac{1}{6\,v^2} 
\langle \Omega | \bar{q} \hat{v} g_s \sigma {\cal G} q | \Omega \rangle \epsilon_{\alpha \mu \nu \sigma} v^\sigma\,,
\label{eq:projection1}
\end{eqnarray}

\begin{eqnarray}
&& \langle \Omega | \bar{q} g_s \sigma_{\alpha\beta} {\cal G}_{\mu\nu} q | \Omega \rangle
= \langle \Omega | \bar{q} g_s \sigma {\cal G} q | \Omega \rangle
\frac{1}{12} \left( g_{\alpha\mu} g_{\beta\nu} - g_{\alpha\nu} g_{\beta\mu} \right)
\nonumber\\
&& + \left[ \frac{1}{12} \langle \Omega | \bar{q} g_s \sigma {\cal G} q | \Omega \rangle
- i \frac{2m_q}{3} \langle \Omega | \bar{q} \hat{v} \frac{(vD)}{v^2} q | \Omega \rangle
-  \frac{2}{3} \langle \Omega | \bar{q} \frac{(vD)^2}{v^2} q | \Omega \rangle \right] {\rm S}_{\alpha\beta\mu\nu} \,.
\nonumber\\
\label{eq:projection2}
\end{eqnarray}

\noindent
We have explicitly separated the medium specific contributions from the vacuum projections. 
Medium specific contributions are either condensates that contain the medium four-velocity $v$ or combinations of 
condensates that appear in vacuum and medium. The latter ones are always written with angled brackets. Applying vacuum 
projections to the medium specific terms makes them zero in the vacuum limit.

In order to give an example for this procedure we briefly proof Eq.~(\ref{eq:projection1}). Due to Lorentz covariance we 
write
\begin{eqnarray}
\langle \Omega | \bar{q} \gamma_5 \gamma_\alpha {\cal G}_{\mu\nu} q | \Omega \rangle &=& 
A \, \epsilon_{\alpha\mu\nu\sigma} v^\sigma\,.
\label{eq:ansatz_projection}
\end{eqnarray}

\noindent
Our aim is to determine the Lorentz scalar $A$. We can write for any four-vector $v_\mu$
\begin{eqnarray}
\hat{v} \sigma^{\mu\nu} {\cal G}_{\mu\nu} &=& 
i \, v_\alpha \gamma^\alpha \gamma^\mu \gamma^\nu {\cal G}_{\mu\nu} \,.
\end{eqnarray}

\noindent
Expanding the product of Dirac matrices in terms of the Clifford algebra, one obtains
\begin{eqnarray}
\gamma^\alpha \gamma^\mu \gamma^\nu &=& 
g^{\mu\nu} \gamma^\alpha + g^{\nu\alpha} \gamma^\mu + g^{\alpha\mu} \gamma^\nu 
+ i \epsilon^{\sigma \alpha \mu \nu} \gamma_5 \gamma_\sigma \,.
\end{eqnarray}

\noindent
Due to the equations of motion and by the definition of the gluon field strength tensor, one can show that
\begin{eqnarray}
\langle \Omega | \bar{d} v^\mu \gamma^\nu {\cal G}_{\mu\nu} d | \Omega \rangle &=& 0 \,.
\end{eqnarray}

\noindent
Altogether, this gives the relation
\begin{eqnarray}
\langle \Omega | \bar{d} \hat{v} \sigma^{\mu\nu} {\cal G}_{\mu\nu} d | \Omega \rangle &=& 
- \langle \Omega | \bar{d} \gamma_5 \gamma_\sigma {\cal G}_{\mu\nu} d | \Omega \rangle 
\epsilon^{\sigma \alpha \mu \nu} v_\alpha \,.
\label{arbitrary_fourvector}
\end{eqnarray}

\noindent
Contracting (\ref{eq:ansatz_projection}) with $\epsilon^{\alpha \mu \nu \sigma}$ and using 
$\epsilon^{\alpha \mu \nu \sigma} \epsilon_{\alpha \mu \nu}^{\phantom{\alpha \mu \nu} \tau} = 6 g^{\sigma \tau}$, one can 
show the desired relation (\ref{eq:projection1}).

Let us define the operator $\hat{T}$ in such a way that it creates a traceless expression with respect to Lorentz indices. 
For two Lorentz indices $\hat{T}$ reads
\begin{eqnarray}
\hat{T} \left( {\cal O}_{\mu\nu} \right) &=& {\cal O}_{\mu\nu} - \frac{g_{\mu\nu}}{4} {\cal O}_\alpha^\alpha\,,
\label{eq:trace_op_def}
\end{eqnarray}

\noindent
and we immediately observe that the medium specific terms in (\ref{projection2a}) and in (\ref{projection3a}) originate 
from the contraction of $v^\mu v^\nu /v^2$ with a certain traceless expression:

\begin{eqnarray}
\frac{v^\mu v^\nu}{v^2} \langle \Omega | \bar{q} \, \hat{T} \left( \gamma_\mu \overrightarrow{D}_\nu \right) q | \Omega \rangle 
&=& \left[ i \frac{m_q}{4} \langle \Omega | \bar{q} q | \Omega \rangle 
+ \langle \Omega | \bar{q} \hat{v} \frac{(v\overrightarrow{D})}{v^2} q | \Omega \rangle \right] \,,
\\
\frac{v^\mu v^\nu}{v^2} \langle \Omega | \bar{q} \, \hat{T} \left( \overrightarrow{D}_\mu \overrightarrow{D}_\nu \right) q | \Omega \rangle &=&
\bigg[ \frac{m_q^2}{4} \langle \Omega | \bar{q} q | \Omega \rangle 
- \frac{1}{8} \langle \Omega | \bar{q} g_s \sigma {\cal G} q | \Omega \rangle
\nonumber\\
&& + \langle \Omega | \bar{q} \frac{(v\overrightarrow{D})^2}{v^2} q | \Omega \rangle \bigg] \,.
\end{eqnarray}

\noindent
This is clear since the additional medium contributions can be obtained by performing the complete projection of the 
structure to be projected minus the vacuum projection, which is proportional to products of the metric tensor, 
${\cal O}_{\mu\nu} - \frac{g_{\mu\nu}}{4} g^{\alpha\beta} {\cal O}_{\alpha\beta}$. Therefore, the additional medium 
contributions must be traceless. The generalization to arbitrary Lorentz indices is obvious.

\end{appendix}

%================================================

\end{document}